\documentclass{book}
\usepackage{makeidx,epsfig}
\usepackage{setspace,graphicx}
\usepackage{amssymb}
\usepackage{Generic}
\makeindex

\usepackage{color}

\begin{document}
\title{Dynamical Tunneling --- Theory and Experiment}
\author{Srihari Keshavamurthy and Peter Schlagheck (eds)}

\pagenumbering{roman}
\newcommand{\Authors}[1]{\section*{#1}}
\newcommand{\Affiliations}[1]{#1}
\mainmatter
\pagenumbering{arabic}
\setcounter{chapter}{7}
\chapter[Resonance-assisted tunneling]
        {Resonance-assisted tunneling in mixed regular-chaotic systems}

\Authors{Peter Schlagheck$^1$, Amaury Mouchet$^2$, and Denis Ullmo$^3$}
\Affiliations{$^1$ D\'epartement de Physique, Universit\'e de Li\`ege, 
  4000 Li\`ege, Belgium \\
$^2$ Laboratoire de Math\'ematiques et de Physique Th\'eorique, 
Universit\'e Fran\c{c}ois Rabelais de Tours --- \textsc{cnrs (umr 6083)}, 
F\'ed\'eration Denis Poisson, Parc de Grandmont, 37200 Tours, France \\
$^3$ LPTMS UMR 8626, Univ. Paris-Sud, CNRS, 91405 Orsay Cedex, France}

\section{Introduction}

\subsection{Tunneling in integrable systems}

Since the early days of quantum mechanics, tunneling has been
recognized as one of the hallmarks of the wave character of
microscopic physics.  The possibility of a quantum particle to
penetrate an energetic barrier represents certainly one of the most
spectacular implications of quantum theory and has lead to various
applications in nuclear, atomic and molecular physics as well as in
mesoscopic science.  Typical scenarios in which tunneling manifests
are the escape of a quantum particle from a quasi-bounded region,
the transition between two or more symmetry-related, but classically
disconnected wells (which we shall focus on in the following), as well
as scattering or transport through potential barriers.  The spectrum
of scenarios becomes even richer when the concept of tunneling is
generalized to any kind of classically forbidden transitions in phase
space, i.e.\ to transitions that are not necessarily inhibited by
static potential barriers but by some other constraints of the
underlying classical dynamics (such as integrals of motion).  Such
``dynamical tunneling'' processes arise frequently in molecular
systems \cite{DavHel81JCP} and were realized with cold atoms
propagating in periodically modulated optical lattices
\cite{HenO01N,SteOskRai01S,MouO01PRE}. 
Moreover, the electromagnetic analog of dynamical tunneling was also 
obtained with microwaves in billiards \cite{DemO00PRL}.

Despite its genuinely quantal nature, tunneling is strongly influenced 
by the structure of the underlying classical phase space
(see Ref.~\cite{Cre98} for a review).
This is best illustrated within the textbook example of a one-dimensional
symmetric double-well potential.
In this simple case, the eigenvalue problem can be straightforwardly solved
with the standard Jeffreys-Wentzel-Kramers-Brillouin 
(JWKB) ansatz \cite{LanLifQ}.
The eigenstates of this system are, below the barrier height, obtained by the
symmetric and antisymmetric linear combination of the local ``quasi-modes''
(i.e., of the wave functions that are semiclassically constructed on the
quantized orbits within each well, without taking into account the classically
forbidden coupling between the wells), and the splitting of their energies is
given by an expression of the form
\begin{equation}
  \Delta E = \frac{\hbar \Omega}{\pi} \exp\left[ - \frac{1}{\hbar}
    \int \sqrt{2m(V(x) - E)} dx 
  \right] \, . \label{eq:1dsplit}
\end{equation}
Here $E$ is the mean energy of the doublet, $V(x)$ represents the double well
potential, $m$ is the mass of the particle, $\Omega$ denotes the oscillation
frequency within each well, and the integral in the exponent is performed over
the whole classically forbidden domain, i.e.\ between the inner turning points
of the orbits in the two wells.
Preparing the initial state as one of the quasi-modes (i.e., as the
even or odd superposition of the symmetric and the antisymmetric eigenstate),
the system will undergo Rabi oscillations between the wells with the frequency
$\Delta E / \hbar$.
The ``tunneling rate'' of this system is therefore given by the splitting
(\ref{eq:1dsplit}).
Keeping all classical parameters fixed, it decreases
exponentially with $1 / \hbar$, and, in that sense, one can say
that tunneling ``vanishes'' in the classical limit.

\subsection{Chaos-assisted tunneling}
\label{sec:cat}

The approach presented in the previous section
can be generalized to multidimensional, even non-separable
systems, as long as their classical dynamics is still integrable
\cite{Cre94JPA}.
It breaks down, however, as soon as a non-integrable perturbation is added to
the system, e.g.\ if the one-dimensional double-well potential is exposed to 
a driving that is periodic in time (with period~$\tau$, say).
In that case, the classical phase space of the system generally becomes 
a mixture of both regular and chaotic structures.

As visualized by the stroboscopic Poincar\'e section 
--- which is obtained by retaining the phase space coordinates at 
every integer multiple of the driving period $\tau$ ---
the phase space typically displays two prominent regions of regular motion,
corresponding to the weakly perturbed dynamics within the two wells,
and a small (or, for stronger perturbations, large) layer of chaotic
dynamics that separates the two regular islands from each other.
Numerical calculations of model systems 
in the early nineties \cite{LinBal90PRL,BohO93NP} have shown that the
tunnel splittings in such mixed systems generally become strongly
enhanced compared to the integrable limit.  Moreover, they do no
longer follow a smooth exponential scaling with $1 / \hbar$ as
expressed by Eq.~(\ref{eq:1dsplit}), but display huge, quasi-erratic
fluctuations when $\hbar$ or any other parameter of the system vary  
\cite{LinBal90PRL,BohO93NP}.

\begin{figure}[ht]
  \begin{center}
    \includegraphics*[width=10cm]{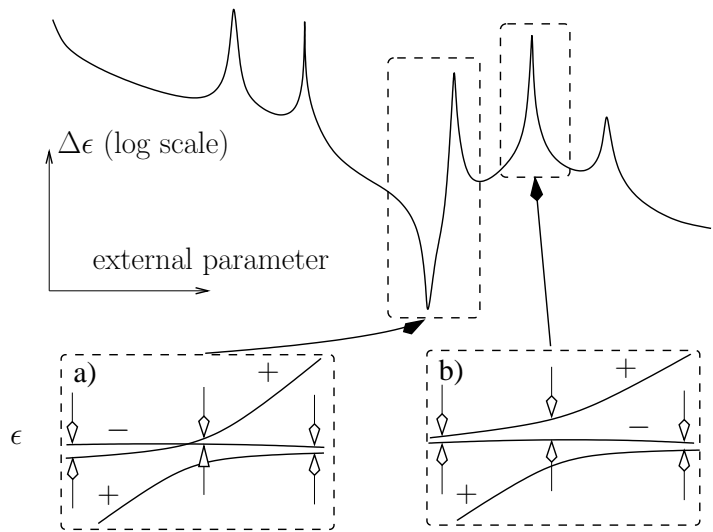}
  \end{center}
  \caption{The two elementary scenarios of huge enhancement (case b)
    or cancellation (case a) of 
    the tunneling splitting between symmetric ($+$) and anti-symmetric 
    ($-$) states
    can be easily understood from the resonant crossing of a third level
    (here a symmetric one)  
    and the corresponding 
    level repulsion between states of the same symmetry class. The
    external parameter that triggers the  
    fluctuation of tunneling can be of quantum origin  (effective
    $\hbar$) or classical. \label{fig:croisement3niveaux}}
\end{figure}

These phenomena are traced back to the specific role that {\em chaotic}
states play in such systems
\cite{BohTomUll93PR,TomUll94PRE,DorFri95PRL,FriDor98PRE}.
In contrast to the integrable case, the tunnel doublets of the localized
quasi-modes are, in a mixed regular-chaotic system, no longer isolated in the
spectrum, but resonantly interact with states that are associated with the
chaotic part of phase space. Due to their delocalized nature
in phase space\footnote{We assume here that
the effects of dynamical localization observed in 
Ref.~\cite{Ishikawa+09a} remain irrelevant.},
such chaotic states typically exhibit a significant overlap with the
boundary regions of both regular wells. Therefore, they may 
provide an efficient coupling mechanism between the quasi-modes -- which becomes
particularly effective whenever one of the chaotic levels is shifted
exactly on resonance with the tunnel doublet.  As illustrated in
Fig.~\ref{fig:croisement3niveaux}, this coupling mechanism generally
enhances the tunneling rate, but may also lead to a complete
suppression thereof, arising at specific values of $\hbar$ or other
parameters \cite{GroO91PRL}.

\begin{figure}[ht]
  \begin{center}
    \includegraphics*[width=10cm]{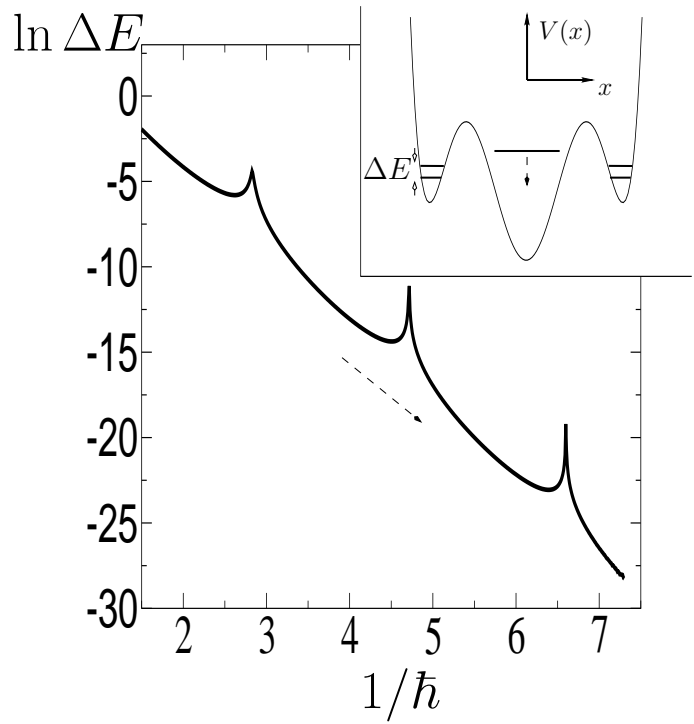}
  \end{center}
  \caption{Resonant tunneling at work for a one dimensional  triple
    well potential (here $V(x)=(x^2-a^2)^2(x^2-b^2)$ 
    with $a=1.75$ and $b=.5$). When $1/\hbar$ is increased, tunneling
    between the two lateral wells is enhanced 
    by several orders of magnitude each time a level in the central well
    crosses down the doublet. 
    \label{fig:DeltaEq6}
  }
\end{figure}

We point out that this type of resonant tunneling does not necessarily 
require the presence of classical chaos and may appear also in integrable
systems, for instance in a one-dimensional symmetric triple-well potential.
This is illustrated in Fig.~\ref{fig:DeltaEq6}, which shows the scaling
of level splittings associated with the two lateral wells as a function of
$1/\hbar$ for such a triple-well potential.
On top of an exponential decrease according to Eq.~(\ref{eq:1dsplit}), 
the splittings display strong spikes occuring whenever the energy 
of a state localized in the central well becomes quasi-degenerate with 
the energies of the states in the two lateral wells.

For mixed regular-chaotic systems, the
validity of this ``chaos-assisted'' tunneling picture was essentially
confirmed by
successfully modeling the chaotic part of the quantum dynamics with
a random matrix from the Gaussian orthogonal ensemble (GOE)
\cite{BohTomUll93PR,TomUll94PRE,LeyUll96JPA}. Using the fact that
the coupling coefficients between the regular states and the
chaotic domain are small, this random matrix ansatz yields a truncated Cauchy
distribution for the probability density to obtain a level splitting of the
size $\Delta E$.
Such a distribution is indeed encountered in the exact quantum splittings,
which was demonstrated for the two-dimensional quartic oscillator
\cite{LeyUll96JPA} as well as, later on, for the driven pendulum Hamiltonian
that describes the tunneling process of cold atoms in periodically modulated
optical lattices \cite{MouO01PRE,MouDel03PRE}.
A quantitative prediction of the {\em average} tunneling rate, however, 
was not possible in the above-mentioned theoretical works.  
As we shall argue later on, this average tunneling rate is directly 
connected to the coupling matrix element between the regular and the 
chaotic states, and the strength of this matrix element was unknown 
and introduced in an ad-hoc way. 

A first step towards this latter problem was undertaken by
Podolskiy and Narimanov \cite{PodNar03PRL} who proposed an
explicit semiclassical expression for the mean tunneling rate in a
mixed system by assuming a perfectly clean, harmonic-oscillator like
dynamics within the regular island and a structureless chaotic sea
outside the outermost invariant torus of the island.  This expression
turned out to be successful for the reproduction of the level
splittings between near-degenerate optical modes that are associated
with a pair of symmetric regular islands in a non-integrable
micro-cavity \cite{PodNar03PRL} (see also Ref.~\cite{PodNar05OL}).
The application to dynamical tunneling in periodically modulated
optical lattices \cite{PodNar03PRL}, for which splittings between the
left- and the right-moving stable eigenmodes were calculated in
Ref.~\cite{MouO01PRE}, seems convincing for low and moderate values of
$1 / \hbar$, but reveals deviations deeper in the semiclassical regime
where plateau structures arise in the tunneling rates.  Further, and
more severe, deviations were encountered in the application of this
approach to tunneling processes in other model systems \cite{BaeKet}.

B\"acker, Ketzmerick, L\"ock, and coworkers \cite{BaeO08PRL,BaeO208PRL}
recently undertook the effort to derive more rigorously the
regular-to-chaotic coupling rate governing chaos-assisted tunneling.
Their approach is based on the construction of an integrable 
approximation for the nonintegrable system, designed to accurately 
describe the motion within the regular islands under consideration.
The coupling rate to the chaotic domain is then determined through 
the computation of matrix elements of the Hamiltonian of the system
within the eigenbasis of this integrable approximation \cite{BaeO08PRL}.
This results in a smooth exponential-like decay of the average tunneling
rate with $1 / \hbar$, which was indeed found to be in very good agreement 
with the exact tunneling rates for quantum maps and billiards
\cite{BaeO08PRL,BaeO208PRL}.
Those systems, however, were designed such as to yield a ``clean'' 
mixed regular-chaotic phase space, containing a regular island and 
a chaotic region which both do not exhibit appreciable substructures 
\cite{BaeO08PRL,BaeO208PRL}.

\subsection{The role of nonlinear resonances}

In more generic systems, such as the quantum kicked rotor or the
driven pendulum \cite{MouO01PRE}, however, even the ``average''
tunneling rates do not exhibit a smooth monotonous behaviour with
$1 / \hbar$, but display peaks and plateau structures that cannot be
accounted for by the above approaches.  To understand the origin of
such plateaus, it is instructive to step back to the conceptually
simpler case of {\em nearly integrable} dynamics, where the
perturbation from the integrable Hamiltonian is sufficiently small
such that macroscopically large chaotic layers are not yet developed 
in the Poincar\'e surface of section.  In such systems,
the main classical phase space features due to the perturbation
consist in chain-like substructures that surround stable periodic orbits 
or equilibrium points of the classical motion. Those substructures
come from \emph{nonlinear resonances} between the internal degrees of 
freedom of the system or, for driven systems, between the 
external driving and the unperturbed oscillations around the central orbit. 
In a similar way as for the quantum pendulum Hamiltonian, such resonances
induce additional tunneling paths in the phase space, which lead to
couplings between states that are located 
near the {\em same} stable orbit \cite{Ozo84JPC,UzeNoiMar83JCP}.

The relevance of this effect for the near-integrable tunneling process
between two symmetry-related wells was first pointed out by Bonci et
al.~\cite{BonO98PRE} who argued that such resonances may lead
to a strong enhancement of the tunneling rate, due to couplings
between lowly and highly excited states within the well which are
permitted by near-degeneracies in the spectrum
In Refs.~\cite{BroSchUll01PRL,BroSchUll02AP}, a
quantitative semiclassical theory of near-integrable tunneling was
formulated on the basis of this principal mechanism.  This theory
allows one to reproduce the exact quantum splittings from purely
classical quantities and takes into account high-order effects such as
the coupling via a sequence of different resonance chains
\cite{BroSchUll01PRL,BroSchUll02AP}.  More recent studies by
Keshavamurthy on classically forbidden coupling processes in
model systems that mimic the dynamics of simple molecules
confirm that the resonance-assisted tunneling scenario prevails not
only in one-dimensional systems that are subject to a periodic driving
(such as the kicked Harper model studied in
Ref.~\cite{BroSchUll01PRL,BroSchUll02AP}), but also in autonomous
systems with two and even three degrees of freedom
\cite{Kes05JCP,Kes05PRE}.

In Refs.~\cite{EltSch05PRL,SchEltUll06,MouEltSch06PRE,WimO06PRL}
resonance-assisted couplings were incorporated in an
approximate manner into the framework of chaos-assisted tunneling
in order to provide a quantitative theory for the regular-to-chaotic
coupling rate.  In this context, it is assumed that the dominant
coupling between regular states within and chaotic states 
outside the island is
provided by the presence of a nonlinear resonance within the island.
A straightforward implementation of this idea yields good agreement
with the exact tunneling rates as far as their average decay with
$1 / \hbar$ in the deep semiclassical limit is concerned.  Moreover,
individual plateaus and peak structures could be traced back to the
influence of specific nonlinear resonances, not only for
double-well-type tunneling in closed or periodic systems
\cite{EltSch05PRL,SchEltUll06,MouEltSch06PRE}, but also for
tunneling-induced decay in open systems \cite{WimO06PRL}.  However,
the predictive power of this method was still rather limited,
insofar as individual tunneling rates at given system parameters
could be over- or underestimated by many orders of magnitude.  In
particular, resonance-assisted tunneling seemed inapplicable in the
``quantum'' limit of large $\hbar$, where direct regular-to chaotic
tunneling proved successful \cite{BaeO08PRL,BaeO208PRL}.

A major advance in this context was achieved by improving the
semiclassical evaluation of resonance-induced coupling processes in
mixed systems, and by combining it with ``direct'' regular-to-chaotic
tunneling \cite{LoeO10PRL}.  This combination resulted, for the first
time, in a semiclassical prediction of tunneling rates in generic
mixed regular-chaotic systems that can be compared with the exact
quantum rates on the level of individual peak structures
\cite{LoeO10PRL}.  This confirms the expectation that nonlinear
resonances do indeed form the ``backbone'' behind non-monotonous
substructures in tunneling rates.  It furthermore suggests that those
rates could, also in systems with more degrees of freedom, possibly be
estimated in a quantitatively satisfactory manner via simple classical
computations, based on the most prominent nonlinear resonances that
are manifested within the regular island.

It is in the spirit of this latter expectation that this contribution
has been written.  Our aim is not to formulate a formal semiclassical
theory of tunneling in mixed systems, which still represents an open
problem that would rather have to be solved on the basis of complex
classical orbits \cite{ShuIke95PRL,ShuIke96PRL,ShuIshIke02JPA}.
Instead, we want to provide a simple, easy-to-implement, yet effective
prescription how to compute the rates and time scales associated with
 tunneling processes solely on the
basis of the classical dynamics of the system, without performing any
diagonalization (not even any application) of the quantum Hamiltonian
or of the time evolution operator.  
This prescription is based on chaos- and
resonance-assisted tunneling in its improved form \cite{LoeO10PRL}.
The main part of this contribution is therefore devoted 
to a detailed description of resonance-assisted tunneling and 
its combination with chaos-assisted tunneling
in the sections \ref{sec:rat} and \ref{sec:combo}, respectively.
We present in section~\ref{sec:kr} the
application of this method to tunneling processes in the quantum
kicked rotor, and discuss possible limitations and future prospects in
the conclusion in section~\ref{sec:cc}.

\section{Theory of resonance-assisted tunneling}

\label{sec:rat}

\subsection{Secular perturbation theory}

\label{sec:sec}

For our study, we restrict ourselves to systems with one degree of freedom 
that evolve under a periodically time-dependent Hamiltonian 
$H(p,q,t) = H(p,q,t + \tau)$.
We suppose that, for a suitable choice of parameters, the classical phase 
space of $H$ is mixed regular-chaotic and exhibits two symmetry-related
regular islands that are embedded within the chaotic sea.
This phase space structure is most conveniently visualized by a stroboscopic
Poincar\'e section, where $p$ and $q$ are plotted at the times 
$t = n \tau (n \in \mathbb{Z})$.
Such a Poincar\'e section typically reveals the presence of chain-like
substructures within the regular islands, which arise due to 
nonlinear resonances between the external driving and the internal
oscillation around the island's center.
Before considering the general situation for which many resonances may
come into play in the tunneling process, we start with the simpler
case where the two islands exhibit a prominent $r$:$s$ 
resonance, i.e., a nonlinear resonance where $s$ internal oscillation 
periods match $r$ driving periods and $r$ sub-islands are visible in 
the stroboscopic section.

The classical motion in the vicinity of the $r$:$s$ resonance is
approximately integrated by secular perturbation theory \cite{LicLie}
(see also Ref.~\cite{BroSchUll02AP}).  For this purpose, we formally
introduce a time-independent Hamiltonian $H_0(p,q)$ that approximately
reproduces the regular motion in the islands and preserves the
discrete symmetry of $H$.  In some circumstances, as
for instance if $H$ is in the nearly integrable regime, $H_0(p,q)$ can
be explicitly computed within some approximation scheme (using for
instance the Lie transformation method \cite{LicLie}).  We stress
though that this will not always be necessary.  Assuming 
the existence of such a $H_0$,
 the phase space generated by this integrable
Hamiltonian consequently exhibits two symmetric wells that are
separated by a dynamical barrier and ``embed'' the two islands of $H$.  
In terms of the action-angle variables $(I,\theta)$ describing the dynamics 
within each of the wells, the total Hamiltonian can be written as
\begin{equation}
  H(I,\theta,t) = H_0(I) + V(I,\theta,t) \label{eq:H}
\end{equation}
where $V$ would represent a weak perturbation in the center of the island
\footnote{In order not to overload the notation, we use the same symbol 
  $H$ for the
  Hamiltonian in the original phase-space variables $(p,q)$ and in the
  action-angle variables $(I,\theta)$.}

The nonlinear $r$:$s$ resonance occurs at the action variable $I_{r:s}$ that
satisfies the condition
\begin{equation}
  r \Omega_{r:s} = s \omega \label{eq:r:s}
\end{equation}
with $\omega = 2 \pi / \tau$ and
\begin{equation}
  \Omega_{r:s} \equiv \left. \frac{d H_0}{d I}\right|_{I=I_{r:s}} \, .
\end{equation}
We now perform a canonical transformation to the frame that corotates with
this resonance.
This is done by leaving $I$ invariant and modifying $\theta$ according to
\begin{equation}
  \theta \mapsto \vartheta = \theta - \Omega_{r:s} t \, .
\end{equation}
This time-dependent shift is accompanied by the transformation
$H \mapsto \mathcal{H} = H - \Omega_{r:s} I$ in order to ensure that 
the new corotating angle variable $\vartheta$ is conjugate to $I$.
The motion of $I$ and $\vartheta$ is therefore described by the new Hamiltonian
\begin{equation}
  \mathcal{H}(I,\vartheta,t) = \mathcal{H}_0(I) + \mathcal{V}(I,\vartheta,t)
\end{equation}
with
\begin{eqnarray}
  \mathcal{H}_0(I) & = & H_0(I) - \Omega_{r:s} I \, , \label{eq:Hrot} \\
  \mathcal{V}(I,\vartheta,t) & = & V(I,\vartheta + \Omega_{r:s} t,t) \, . 
  \label{eq:Vrot}
\end{eqnarray}

The expansion of $\mathcal{H}_0$ in powers of $I - I_{r:s}$ yields
\begin{equation}
  \mathcal{H}_0(I) \simeq \mathcal{H}_0^{(0)} + 
  \frac{(I - I_{r:s})^2}{2 m_{r:s}} +
  \mathcal{O}\left[(I - I_{r:s})^3\right] \label{eq:H0res}
\end{equation}
with a constant 
$\mathcal{H}_0^{(0)} \equiv H_0(I_{r:s}) - \Omega_{r:s} I_{r:s} $ and 
a quadratic term that is characterized by the effective ``mass'' 
parameter $m_{r:s} \equiv [d^2 H_0/d I^2(I_{r:s})]^{-1}$.
Hence, $d \mathcal{H}_0 / d I$ is comparatively small for 
$I \simeq I_{r:s}$, which implies that the co-rotating angle 
$\vartheta$ varies slowly in time near the resonance.
This justifies the application of adiabatic perturbation theory \cite{LicLie},
which effectively amounts, in first order, to replacing 
$\mathcal{V}(I,\vartheta,t)$
by its time average over $r$ periods of the driving (using the fact that 
$\mathcal{V}$ is periodic in $t$ with the period $r \tau$) 
\footnote{This step involves, strictly speaking, another 
  time-dependent canonical
  transformation $(I,\vartheta) \mapsto (\widetilde{I},\widetilde{\vartheta})$
  which slightly modifies $I$ and $\vartheta$ (see also
  Ref.~\cite{BroSchUll02AP}).}.
We therefore obtain, after this transformation, the time-independent
Hamiltonian 
\begin{equation} \label{eq:adiabatic}
  \mathcal{H}(I,\vartheta) = \mathcal{H}_0(I) + \mathcal{V}(I,\vartheta)
\end{equation}
with
\begin{equation}
  \mathcal{V}(I,\vartheta) \equiv 
  \frac{1}{r\tau}\int_0^{r\tau} \mathcal{V}(I,\vartheta,t) dt \, .
\end{equation}

By expanding $V(I,\theta,t)$ in a Fourier series
in both $\theta$ and $t$, i.e. 
\begin{equation}
  V(I,\theta,t) = \sum_{l,m = - \infty}^{\infty} V_{l,m}(I) 
  e^{i l \theta} e^{i m \omega t}
  \label{eq:Vseries}
\end{equation}
with $V_{l,m}(I) = [V_{-l,-m}(I)]^*$, one can straightforwardly 
derive
\begin{equation}
  \mathcal{V}(I,\vartheta) = V_{0,0}(I) +
  \sum_{k=1}^{\infty} 2 V_k(I)\cos(k r \vartheta + \phi_k) \label{eq:Vres}
\end{equation}
defining
\begin{equation}
  V_k(I) e^{i \phi_k} \equiv V_{rk, -sk}(I) \, , \label{eq:Vk}
\end{equation}
i.e., the resulting time-independent perturbation term is ($2\pi/r$)-periodic 
in $\vartheta$.

For the sake of clarity, we start discussing the resulting effective
Hamiltonian neglecting the action dependence of the Fourier
coefficients of $\mathcal{V}(I,\vartheta)$.  We stress
that this dependence can be implemented in a relatively straightforward
way using Birkhoff-Gustavson normal-form coordinates (cf
section~\ref{sec:ac} below); it is actually important to obtain a
good quantitative accuracy.  For now, however, we replace $V_k(I)$ by
$V_k \equiv V_k(I=I_{r:s})$ in Eq.~(\ref{eq:Vres}).  Neglecting
furthermore the term $V_{0,0}(I)$, we obtain the effective integrable
Hamiltonian 
\begin{equation}
  H_{\rm res}(I,\vartheta) = H_0(I) - \Omega_{r:s} I + 
  \sum_{k=1}^\infty 2 V_k \cos(k r \vartheta + \phi_k) \label{heff}
\end{equation}
for the description of the classical dynamics in the vicinity of the
resonance.  We shall see in
section~\ref{sec:pend} that the parameters of $H_{\rm res}$ relevant to
the tunneling process can be extracted directly from the classical
dynamics of $H(t)$, which is making Eq.~(\ref{heff}) particularly
valuable.

\subsection{The pendulum approximation}

\label{sec:pend}

The quantum implications due to the presence of this nonlinear resonance can 
be straightforwardly inferred from the direct semiclassical quantization
of $H_{\rm res}$,  given by
\begin{equation}
  \hat{H}_{\rm  res} = H_0(\hat{I}) - \Omega_{r:s} \hat{I} + 
  \sum_{k=1}^\infty 2 V_k \cos(k r \hat{\vartheta} + \phi_k) \, . \label{eq:heffqm}
\end{equation}
Here we introduce the action operator 
$\hat{I} \equiv - i \hbar \partial / \partial \vartheta$
and assume anti-periodic boundary conditions in $\vartheta$ in order to 
properly account for the Maslov index in the original phase space 
\cite{Ozo84JPC}.
In accordance with our assumption that the effect of the resonance is 
rather weak, we can now apply quantum perturbation theory to the 
Hamiltonian (\ref{eq:heffqm}), treating the $\hat{I}$-dependent 
``kinetic'' terms as unperturbed part
and the $\hat{\vartheta}$-dependent series as perturbation.
The unperturbed eigenstates are then given by the (anti-periodic) 
eigenfunctions 
$\langle\vartheta|n\rangle = (2\pi)^{-1/2} \exp[i (n + 1/2) \vartheta]$ 
($n \geq 0$) of the action operator $\hat{I}$ with the eigenvalues
\begin{equation}
  I_n = \hbar ( n + 1/2) \, .
\end{equation}

As is straightforwardly evaluated, the presence of the perturbation 
induces couplings
between the states $|n\rangle$ and $|n+kr\rangle$ with the matrix elements
\begin{equation}
  \langle n+kr|\hat{H}_{\rm res}|n\rangle = V_k e^{i \phi_k} \label{eq:matrelem}
\end{equation}
for strictly positive integer $k$.
As a consequence, the ``true'' eigenstates $|\psi_n\rangle$ of 
$\hat{H}_{\rm res}$ contain admixtures from unperturbed modes 
$|n'\rangle$ that satisfy the selection rule
$|n' - n| = kr$ with integer $k$.
They are approximated by the expression
\begin{eqnarray}
  |\psi_n\rangle & = & |n\rangle + \sum_{k} 
  \frac{\langle n+kr|\hat{H}_{\rm res}|n\rangle}
  {E_n - E_{n+kr} + k s \hbar \omega} |n+kr\rangle + \nonumber \\
  & & + \sum_{k,k'} 
  \frac{\langle n+kr|\hat{H}_{\rm res}|n+k'r\rangle}
  {E_n - E_{n+kr}+ k s \hbar \omega}
  \frac{\langle n+k'r|\hat{H}_{\rm res}|n\rangle}
  {E_n - E_{n+k'r} + k' s \hbar \omega} 
  |n+kr\rangle + \ldots \label{npert}
\end{eqnarray}
where $E_n \equiv H_0(I_n)$ denote the unperturbed eigenenergies of $H_0$
and  the resonance condition (\ref{eq:r:s}) is used. 
The summations in Eq.~(\ref{npert}) are generally finite due to the
finiteness of the phase space area covered by the regular region.

Within the quadratic approximation of $H_0(I)$ around $I_{r:s}$, we obtain 
from Eqs.~(\ref{eq:Hrot}) and (\ref{eq:H0res})
\begin{equation}
  E_n \simeq H_0(I_{r:s}) + \Omega_{r:s} ( I_n - I_{r:s}) 
  + \frac{1}{2m_{r:s}}(I_n - I_{r:s})^2 \, .
  \label{eq:en}
\end{equation}
This results in the energy differences
\begin{equation}
  E_n - E_{n+kr} + k s \hbar \omega \simeq
  \frac{1}{2m_{r:s}} (I_n - I_{n+kr}) (I_n + I_{n+kr} - 2 I_{r:s})
  \, . \label{Ediff}
\end{equation}
From this expression, we see that the admixture between $|n\rangle$
and $|n'\rangle$ becomes particularly strong if the $r$:$s$ resonance
is symmetrically located between the two tori that are associated with
the actions $I_n$ and $I_{n'}$ --- i.e., if $I_n + I_{n'} \simeq 2
I_{r:s}$.  The presence of a significant nonlinear resonance within a
region of regular motion provides therefore an efficient mechanism to
couple the local ``ground state'' --- i.e, the state that is semiclassically
localized in the center of that region (with action variable $I_0 < I_{r:s}$)
--- to a highly excited state (with action variable $I_{kr} > I_{r:s}$).

It is instructive to realize that the Fourier coefficients $V_k$ of the
perturbation operator decrease rather rapidly with increasing $k$.
Indeed, one can derive under quite general circumstances the asymptotic
scaling law
\begin{equation}
  V_k \sim (kr)^\gamma V_0 \exp[ - k r \Omega_{r:s} t_{\rm im}(I_{r:s}) ] 
  \label{Vk}
\end{equation}
for large $k$, which is based on the presence of singularities of the
complexified tori of the integrable approximation $H_0(I)$
(see Eq.~(66) in Ref.~\cite{BroSchUll02AP}).
Here $t_{\rm im}(I)$ denotes the imaginary time that elapses from the (real) 
torus with action $I$ to the nearest singularity in complex phase space, 
$\gamma$ corresponds to the degree of the singularity, and $V_0$ contains 
information about the corresponding residue near the singularity as well 
as the strength of the perturbation.
The expression (\ref{Vk}) is of little practical relevance as far as the
concrete determination of the coefficients $V_k$ is concerned.
It permits, however, to estimate the relative importance of different
perturbative pathways connecting the states $|n\rangle$ and $|n+kr\rangle$ 
in Eq.~(\ref{npert}).
Comparing e.g.\ the amplitude $\mathcal{A}_2$ associated with a single step
from $|n\rangle$ to $|n+2r\rangle$ via $V_2$ and the amplitude 
$\mathcal{A}_1$ associated with two steps from $|n\rangle$ to 
$|n+2r\rangle$ via $V_1$, we obtain from
Eqs.~(\ref{Ediff}) and (\ref{Vk}) the ratio
\begin{equation}
  \mathcal{A}_2 / \mathcal{A}_1 \simeq  
  \frac{2^{\gamma-1} r^{2-\gamma} \hbar^2}{m_{r:s} V_0} 
  e^{i (\phi_2 - 2 \phi_1)}
\end{equation}
under the assumption that the resonance is symmetrically located in between
the corresponding two tori 
(in which case we would have $I_{n+r} \simeq I_{r:s}$).
Since $V_0$ can be assumed to be finite in mixed regular-chaotic systems, we
infer that the second-order process via the stronger coefficient $V_1$ will
more dominantly contribute to the coupling between $|n\rangle$ and 
$|n+2r\rangle$ in the semiclassical limit $\hbar \to 0$.

A similar result is obtained from a comparison of the one-step process via
$V_k$ with the $k$-step process via $V_1$, where we again find that the 
latter more dominantly contributes to the coupling between $|n\rangle$ and 
$|n+kr\rangle$ in the limit $\hbar \to 0$.
We therefore conclude that in mixed regular-chaotic systems the semiclassical
tunneling process is adequately described by the lowest non-vanishing term 
of the sum over the $V_k$ contributions, which in general is given by 
$V_1 \cos( r \vartheta + \phi_1 )$\footnote{Exceptions from this general 
  rule typically arise in the presence of discrete
  symmetries that, e.g., forbid the formation of resonance chains with an odd
  number of sub-islands and therefore lead to $V_1 = 0$ for an $r$:$s$
  resonance with an odd $r$.}.
Neglecting all higher Fourier components $V_k$ with $k > 1$ and making the 
quadratic approximation of $H_0$ around $I = I_{r:s}$, we finally obtain
an effective pendulum-like Hamiltonian 
\begin{equation}
  H_{\rm res}(I,\vartheta) \simeq \frac{(I - I_{r:s})^2}{2 m_{r:s}} 
  + 2 V_{r:s} \cos( r \vartheta + \phi_1 ) \label{pend}
\end{equation}
with $V_{r:s} \equiv V_1$ \cite{EltSch05PRL}.

\begin{figure}[ht]
  \begin{center}
    \includegraphics*[width=\textwidth]{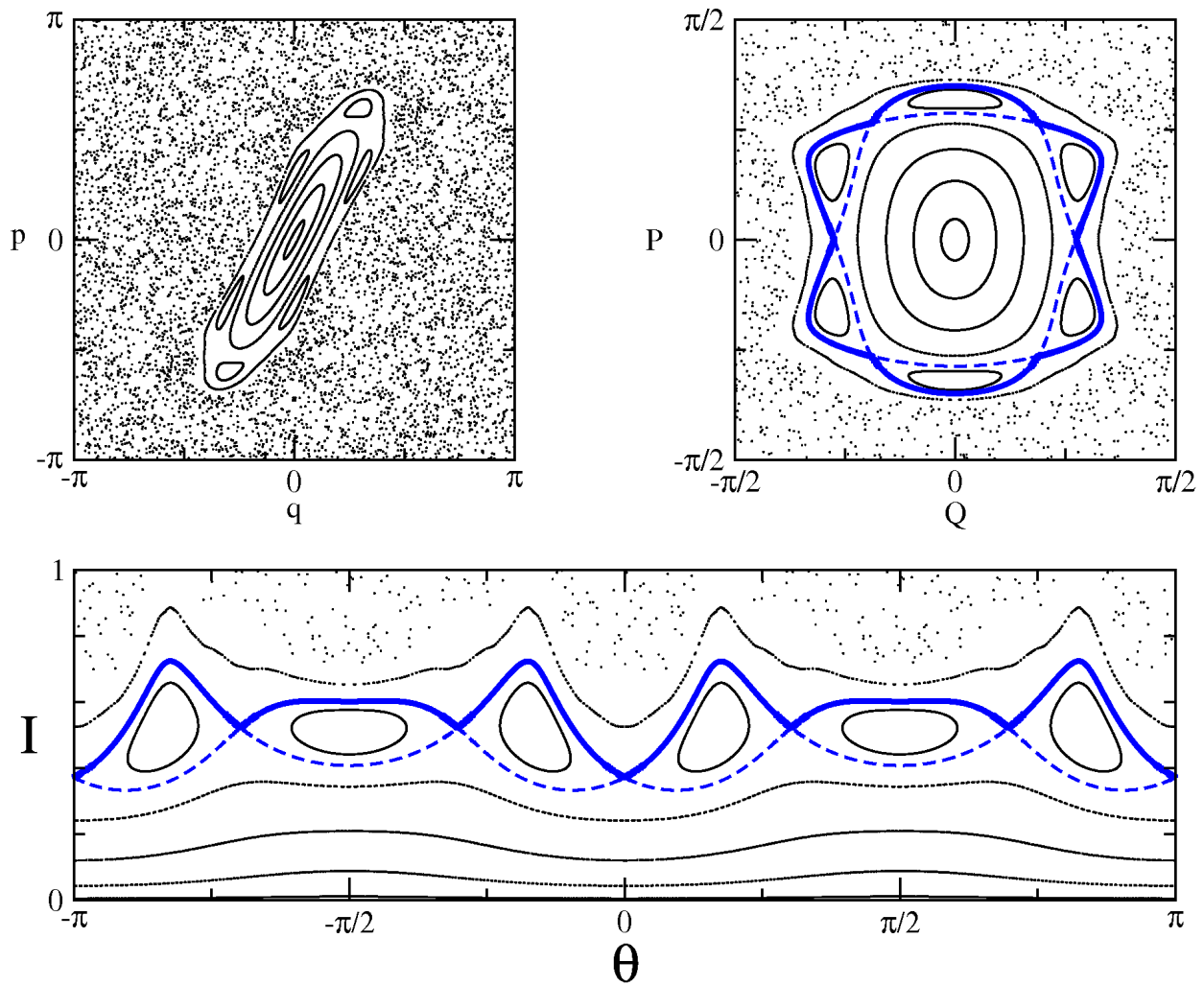}
  \end{center}
  \caption{
    Classical phase space of the kicked rotor Hamiltonian at $K = 3.5$
    showing a regular island with an embedded 6:2 resonance.
    The phase space is plotted in the original $(p,q)$ coordinates 
    (upper left panel), in approximate normal-form coordinates $(P,Q)$ 
    defined by Eqs.~(\ref{eq:BGp}) and (\ref{eq:BGq}) (upper right panel), 
    and in approximate action-angle variables $(I,\vartheta)$ (lower panel).
    The thick solid and dashed lines represent the ``outer'' and ``inner'' 
    separatrix of the resonance, respectively.
    \label{fg:sep}
  }
\end{figure}

This simple form of the effective Hamiltonian allows us to determine
the parameters $I_{r:s}$, $m_{r:s}$ and $V_{r:s}$ from the
Poincar{\'e} map of the classical dynamics, without explicitly using
the transformation to the action-angle variables of $H_0$.  To this
end, we numerically calculate the monodromy matrix $M_{r:s} \equiv
\partial(p_f,q_f) / \partial(p_i,q_i)$ of a stable periodic point of
the resonance (which involves $r$ iterations of the stroboscopic map)
as well as the phase space areas $S_{r:s}^+$ and $S_{r:s}^-$ that are
enclosed by the outer and inner separatrices of the resonance,
respectively (see also Fig.~\ref{fg:sep}).  Using the fact that the
trace of $M_{r:s}$ as well as the phase space areas $S_{r:s}^\pm$
remain invariant under the canonical transformation to
$(I,\vartheta)$, we infer
\begin{eqnarray}
  I_{r:s} & = & \frac{1}{4 \pi} ( S_{r:s}^+ + S_{r:s}^- ) \, , 
  \label{eq:area} \\
  \sqrt{2 m_{r:s} V_{r:s}} & = & \frac{1}{16} ( S_{r:s}^+ - S_{r:s}^- ) \, ,
  \label{eq:sep} \\
  \sqrt{\frac{2 V_{r:s}}{m_{r:s}}} & = & \frac{1}{r^2 \tau} \arccos({\rm tr} 
  \, M_{r:s}/2) \label{eq:trm}
\end{eqnarray}
from the integration of the dynamics generated by $H_{\rm res}$
\cite{TomGriUll95PRL}.  Eqs.~(\ref{eq:area})-(\ref{eq:trm}) make it
possible to derive the final expressions for the tunneling rates
directly from the dynamics of $H(t)$, without explicitly
having to construct the integrable approximation $H_0$ and making the 
Fourier analysis of $V(p,q,t) = H(p,q,t) -H_0(p,q)$. 
As this construction of the integrable approximation
may turn out to be highly non-trivial in the mixed regime,
avoiding this step is actually an essential ingredient to make
the approach we are following practical. We note though that 
improving the quadratic approximation~(\ref{pend}) for
$H_0$ is sometimes necessary, but this does not present any 
fundamental difficulty.

\subsection{Action dependence of the coupling coefficients}

\label{sec:ac}

Up to now, and in our previous publications
\cite{BroSchUll01PRL,BroSchUll02AP,EltSch05PRL,SchEltUll06,MouEltSch06PRE},
we completely neglected the action dependence of the coupling coefficients
$V_k(I)$.
This approximation should be justified in the semiclassical limit of 
extremely small $\hbar$, where resonance-assisted tunneling generally 
involves multiple coupling processes \cite{BroSchUll02AP} and transitions
across individual resonance chains are therefore expected to take place 
in their immediate vicinity in action space.
For finite $\hbar$, however, the replacement $V_k(I) \mapsto V_k(I_{r:s})$,
permitting the direct quantization in action-angle space, is, in general, 
not sufficient to obtain an accurate reproduction of the quantum 
tunneling rates. 
We show now how this can be improved.

To this end, we make the general assumption that the classical Hamiltonian 
$H(p,q,t)$ of our system is analytic in $p$ and $q$ in the vicinity of the
regular islands under consideration.
It is then possible to define an analytical canonical transformation from
$(p,q)$ to Birkhoff-Gustavson normal-form coordinates $(P,Q)$ 
\cite{Bir66,Gus66AJ} that satisfy
\begin{eqnarray}
P & = & - \sqrt{2 I} \sin \theta \, , \label{eq:BGp} \\
Q & = & \sqrt{2 I} \cos \theta \,     \label{eq:BGq}
\end{eqnarray}
and that can be represented in power series in $p$ and $q$.
The ``unperturbed'' integrable Hamiltonian $H_0$ therefore depends only on
$I = (P^2 + Q^2)/2$.

Writing 
\begin{equation}
  e^{\pm i l \theta} = \left( \frac{Q \mp i P}{\sqrt{2I}} \right)^l
\end{equation}
for positive $l$, we obtain, from Eq.~(\ref{eq:Vseries}), the series
\begin{equation}
  V(I,\theta,t) = \sum_{m=-\infty}^{\infty} \left\{ V_{0,m}(I) 
  + \sum_{l=1}^{\infty} \frac{1}{\sqrt{2I}^l}
  \left[ V_{l,m}(I) (Q - i P)^l + V_{-l,m}(I) (Q + i P)^l \right] \right\}
  e^{i m \omega t}
\end{equation}
for the perturbation.  Using the fact that $V(I,\theta,t)$ is analytic
in $P$ and $Q$, we infer that $V_{l,m}(I)$ must scale at least as $I^{l/2}$.
By virtue of (\ref{eq:Vk}), this implies the scaling $V_k(I) \propto I^{rk/2}$
for the Fourier coefficients of the time-independent perturbation term 
that is associated with the $r$:$s$ resonance.
Making the ansatz $V_k(I) \equiv I^{rk/2} \tilde{v}_k$ (and neglecting the 
residual action dependence of $\tilde{v}_k$), we rewrite Eq.~(\ref{eq:Vres}) 
as\footnote{This involves, strictly speaking, another canonical 
transformation of $P$ and $Q$ to the frame that co-rotates with the resonance.}
\begin{equation}
  \mathcal{V}(I,\vartheta) = V_{0,0}(I) +
  \sum_{k=1}^{\infty} \frac{\tilde{v}_k}{2^{kr/2}}
  \left[ (Q - i P)^{kr} e^{i \phi_k} + (Q + i P)^{kr} e^{- i \phi_k} \right] 
  \, . \label{eq:Vres2}
\end{equation}
Each term in the sum is given by the well-known Birkhoff  normal form 
generically describing a~$r\geq3$ resonance
when the bifurcation of the stable periodic orbit is controlled by one
single parameter (see, e.g., Eq.~(4.70) in Chapter 4 of Ref.~\cite{Ozorio88a} 
or Eqs.~(3.3.17) and (3.3.18) in Ref.~\cite{Leboeuf/Mouchet99a} for another
simple derivation of the action dependence of~$V_k(I)$).
The term $V_{0,0}(I)$ is neglected in the following as it does not lead 
to any coupling between different unperturbed eigenstates in the quantum 
system.

One comment is in order here.  The small parameter in the adiabatic
approximation~(\ref{eq:adiabatic}) is the difference 
$(I \! - \! I_{r:s}$), 
while in the derivation of Eq.~(\ref{eq:Vres2}) we have
neglected higher powers of $I$ and thus assumed the action $I$ itself
to be small (strictly speaking we should work near one
bifurcation only).
We thus mix a development near the resonant torus with one
near the center of the island.
This may eventually become problematic if (i) the resonance chain is 
located far away from the center of the island, in which case the 
associated coupling strength may contain a nonnegligible relative error 
when being computed via the assumption $V_k(I) \equiv I^{rk/2} \tilde{v}_k$ 
with constant $\tilde{v}_k$, and if (ii) that coupling strength happens to 
appear rather often in the main perturbative chain that connects the 
quasimodes of the island to the chaotic sea, which generally would be 
the case for low-order resonances with relatively small $r$.\footnote{
Corrections to the form~(\ref{eq:Vres2}) should also arise in the
presence of prominent \emph{secondary} resonances, which occur when primary 
resonances start to overlap and create chains of sub-islands nested inside 
the primary island chains.}
Otherwise, we expect that this inconsistency in the definition of the 
regimes of validity of our perturbative approach does not lead to a 
significant impact on the numerical values of the semiclassical tunneling 
rates, which indeed seems to be confirmed by numerical evidence to be 
discussed below.

This being said, the quantization of the resulting classical Hamiltonian 
can now be carried out 
in terms of the ``harmonic oscillator'' variables $P$ and $Q$ and 
amounts to introducing the standard ladder operators 
$\hat{a}$ and $\hat{a}^\dagger$ according to
\begin{eqnarray}
  \hat{a} & = & \frac{1}{\sqrt{2 \hbar}} ( \hat{Q} + i \hat{P} ) \, , \\
  \hat{a}^\dagger & = & \frac{1}{\sqrt{2 \hbar}} ( \hat{Q} - i \hat{P} ) \, .
\end{eqnarray}
This yields the quantum Hamiltonian 
\begin{equation}
  \hat{H}_{\rm res} = H_0(\hat{I}) - \Omega_{r:s} \hat{I} + 
  \sum_{k=1}^\infty \tilde{v}_k \hbar^{kr/2}
  \left[ \hat{a}^{kr} e^{-i \phi_k} + (\hat{a}^\dagger)^{kr} e^{i
      \phi_k} \right]  
  \label{eq:heffqm2}
\end{equation}
with $\hat{I} \equiv \hbar (\hat{a}^\dagger \hat{a} + 1/2)$.
As for Eq.~(\ref{eq:heffqm}), perturbative couplings are introduced 
only between unperturbed eigenstates $|n\rangle$ and $|n'\rangle$ 
that exhibit the selection rule $|n' - n| = kr$ with integer $k$.
The associated coupling matrix elements are, however, different from 
Eq.~(\ref{eq:matrelem}) and read
\begin{eqnarray}
  \langle n+kr|\hat{H}_{\rm res}|n\rangle & = & \tilde{v}_k
  \sqrt{\hbar}^{kr}  
  e^{i \phi_k} \sqrt{\frac{(n+kr)!}{n!}} \nonumber \\
  & = & V_k(I_{r:s}) e^{i \phi_k} \left(\frac{\hbar}{I_{r:s}}\right)^{kr/2} 
  \sqrt{\frac{(n+kr)!}{n!}} \label{eq:matrelem2}
\end{eqnarray}
for strictly positive $k$.  Close the resonance, i.e. more
formally taking the semiclassical limit $n \to \infty$ keeping $k$ and
$\delta = (I_{r:s}/\hbar \! - \! n)$ fixed, and making use of the
Stirling formula $n!  \simeq \sqrt{2\pi n} (n/e)^n$,
Eq.~(\ref{eq:matrelem2}) reduces to $V_k(I_{r:s}) e^{i \phi_k} $.
The difference becomes, on the other hand,
particularly pronounced if the $r$:$s$ resonance is, in phase space, rather
asymmetrically located in between the invariant tori that correspond
to the states $|n\rangle$ and $|n+kr\rangle$ --- i.e., if $I_{r:s}$ is
rather close to $I_n$ or to $I_{n+kr}$.  In that case,
Eq.~(\ref{eq:matrelem}) may, respectively, strongly over- or
underestimate the coupling strength between these states.

\subsection{Multi-resonance processes}

Up to this point, we considered the couplings between quasi-modes
generated by a given resonance.
In general, however, several of them may play a role for the coupling to
the chaotic sea, giving rise to multi-resonance transitions across subsequent
resonance chains in phase space \cite{BroSchUll01PRL,BroSchUll02AP}.
As was argued in the context of near-integrable systems \cite{BroSchUll02AP},
such multi-resonance processes are indeed expected to dominate over couplings
involving only one single resonance in the deep semiclassical limit
$\hbar \to 0$. 

The description of the coupling process across several consecutive resonances
requires a generalization of Eq.~(\ref{npert}) describing the modified 
eigenstate due to resonance-induced admixtures.
We restrict ourselves, for this purpose, to including only the first-order 
matrix elements $\langle n+r|\hat{H}_{\rm res}^{(r:s)}|n\rangle$ for each
resonance 
[i.e., only the matrix elements with $k=1$ in
Eq.~(\ref{eq:matrelem2})].
For the sake of clarity, we furthermore consider the particular case of 
coupling processes that start in the lowest locally quantized 
eigenmode with node number $n=0$ (the generalization to initial $n
\neq 0$  being straightforward).  The prescription we use is to
consider that, although the approximation (\ref{heff}) is valid
for only one resonance at a time, it is possible to sum the
contributions obtained from different resonances.
Considering a sequence of consecutive $r$:$s$, $r'$:$s'$, $r''$:$s''$ \ldots 
resonances, we obtain in this way
\begin{eqnarray}
  |\psi_0\rangle & = & |0\rangle + \sum_{k>0} \left( \prod_{l=1}^k 
  \frac{\langle lr|\hat{H}_{\rm res}^{(r:s)}|(l-1)r\rangle}
  {E_0 - E_{lr} + l s \hbar \omega} \right) \times \nonumber \\
  & \times & \left\{ |kr\rangle + \sum_{k'>0} \left( \prod_{l'=1}^{k'}
  \frac{\langle kr + l'r' |\hat{H}_{\rm res}^{(r':s')}|kr + (l'-1)r'\rangle}
  {E_0 - E_{kr+l'r'} + (ks +l's') \hbar \omega} \right) \times \right.
  \nonumber \\
  & & \times \left[ |kr + k'r'\rangle + \sum_{k''>0} 
    \left( \prod_{l''=1}^{k''}\frac{\langle kr + k'r' + l''r'' |
      \hat{H}_{\rm res}^{(r'':s'')}|kr  + k'r' + (l''-1)r''\rangle}
    {E_0 - E_{kr+k'r'+l''r''} + (ks +l's'+l''s'') \hbar \omega} \right) 
     \times \right. \nonumber \\
  & & \left. \left. \quad  \times \left( |kr + k'r' + k''r''\rangle + 
    \sum_{k'''>0} \ldots \right) \right] \right\}
\label{eq:multi_amp}
\end{eqnarray}
for the modified ``ground state'' within the island.
Given an excited quasi-mode $n$ far from the interior of the island,
the overlap $\langle n|\psi_0\rangle$ obtained from
Eq.~(\ref{eq:multi_amp}) will in most cases be exponentially dominated
by one or a few contributions.  There is no systematic way to identify
them a priori, although some guiding principle can be used in this
respect \cite{BroSchUll02AP}.

Quite naturally, for instance, low-order resonances, with
comparatively small $r$ and $s$, will, in general, give larger
contribution than high-order resonances with comparable winding
numbers $s/r$ but larger $r$ and $s$, due to the strong differences in
the sizes of the mean coupling matrix elements $V_{r:s}$ [see, e.g.,
Eq.~(\ref{Vk})].  In the same way, sequences of couplings involving
small denominators, i.e.\ energy differences like Eq.~(\ref{Ediff})
that are close to zero, and thus intermediate steps symmetrically
located on each side of a resonance, will tend to give larger
contributions. In the small $\hbar$ limit this will tend to favor
multi-resonance processes. Conversely, for intermediate values of $\hbar$ 
(in terms of the area of the regular region) the main
contributions can be obtained from the lowest-order resonances.  With
few exceptions --- especially concerning low-order resonances that 
are located close to the 
center of the island, thereby leading to relatively 
large energy denominators and small admixtures --- this rule is 
generally observed for
the semiclassical calculation of the eigenphase splittings we shall
consider in section~\ref{sec:kr}.

\section{The combination with chaos-assisted tunneling}
\label{sec:combo}

We now discuss the implication of such nonlinear resonances on the
tunneling process between the two symmetry-related regular islands
under consideration.  In the quantum system, these islands support
(for not too large values of $\hbar$) locally quantized eigenstates or
``quasimodes'' with different node numbers $n$, which, due to the
symmetry, have the same eigenvalues in both islands.  In our case of a
periodically driven system, these eigenvalues can be the eigenphases
$\varphi_n$ of the unitary time evolution (Floquet) operator $\hat{U}$  
over one period $\tau$ of the driving, or, alternatively, 
the quasienergies~$E_n$ such that $\varphi_n = - E_n \tau / \hbar$ 
(modulo $2 \pi$).

The presence of a small (tunneling-induced) coupling between the 
islands lifts the degeneracy of
the eigenvalues and yields the symmetric and antisymmetric linear
combination of the quasimodes in the two islands as ``true''
eigenstates of the system.  A nonvanishing splitting $\Delta \varphi_n
\equiv |\varphi_n^+ - \varphi_n^-|$ consequently arises between the
eigenphases $\varphi_n^\pm$ of the symmetric and the antisymmetric
state, which is related to the splitting $\Delta E_n \equiv |E_n^+ -
E_n^-|$ of the quasi-energies $E_n^\pm$ through $\Delta \varphi_n =
\tau \Delta E_n / \hbar$.

\subsection{Resonance-assisted tunneling in near-integrable systems}

We start by considering a system in the nearly integrable regime.  
In that case, we can assume the presence of a (global) integrable
Hamiltonian $H_0(p,q)$ that describes the dynamics in the entire phase space
to a very good approximation\footnote{Formally, this Hamiltonian is not
identical with the unperturbed approximation $H_0(I)$ introduced in section
\ref{sec:sec} as the definition of the latter is restricted to one well only. 
It is obvious, however, that $H_0(I)$ can be determined from $H_0(p,q)$, 
e.g.\ by means of the Lie transformation method \cite{LicLie}.}.
The energy splittings for the corresponding quantum Hamiltonian
$\hat{H}_0 \equiv H_0(\hat{p},\hat{q})$ 
can be semiclassically calculated 
via an analytic continuation of the invariant tori to the complex domain
\cite{Cre94JPA}.  
This generally yields the splittings
\begin{equation}
  \Delta E_n^{(0)} = \frac{\hbar \Omega_n}{\pi} \exp( - \sigma_n / \hbar ) 
  \label{eq:split0}
\end{equation}
(up to a numerical factor of order one)
 where $\Omega_n$ is the classical oscillation frequency associated with the 
$n$th quantized torus and $\sigma_n$ denotes the imaginary part of the action 
integral along the complex path that joins the two symmetry-related tori.

The main effect of nonlinear resonances in the non-integrable system
is, as was discussed in the previous subsections, to induce
perturbative couplings between quasimodes of different excitation
within the regular islands.  For the nearly integrable systems this
can already lead to a substantial enhancement of the splittings
$\Delta E_n$ as compared to Eq.~(\ref{eq:split0})
\cite{BroSchUll01PRL,BroSchUll02AP}.  
As can be derived within quantum perturbation theory, 
the presence of a prominent $r$:$s$ resonance modifies the
splitting of the local ``ground state'' in the island (i.e., the state with
vanishing node number $n=0$)
according to
\begin{equation}\label{}
  \Delta \varphi_0 = \Delta \varphi_0^{(0)} +
  \sum_{k=1}^{k_c} 
  |\mathcal{A}_{kr}^{(r:s)}|^2  
  \Delta \varphi_{kr}^{(0)} \label{eq:split1}
\end{equation}
(using $\Delta \varphi_{n}^{(0)} \gg \Delta \varphi_0^{(0)}$ for $n > 0$),
where $\mathcal{A}_{kr}^{(r:s)} \equiv \langle kr | \psi_0 \rangle$
denotes the admixture of the $(kr)$th excited unperturbed component $|
kr \rangle$ to the perturbed ground state $| \psi_0 \rangle$ according
to Eq.~(\ref{npert}) [possibly using Eq.~(\ref{eq:matrelem2}) instead
of (\ref{eq:matrelem})]. The maximal number~$k_c$ of coupled states 
is provided by the finite size of the island according to
\begin{equation}\label{eq:kc}
  k_c=\left[\frac{1}{r} \left( 
    \frac{\mbox{area of the island}}{2\pi \hbar} - \frac{1}{2} \right) 
    \right]
\end{equation}
where the bracket stands for the integer part.
The rapid decrease of the amplitudes
$\mathcal{A}_{kr}^{(r:s)}$ with $k$ is compensated by an exponential
increase of the unperturbed splittings $\Delta \varphi_{kr}^{(0)}$,
arising from the fact that the tunnel action $\sigma_n$ in
Eq.~(\ref{eq:split0}) generally decreases with increasing $n$.
The maximal contribution to the modified ground state splitting is
generally provided by the state $| kr \rangle$ for which $I_{kr} + I_0
\simeq 2 I_{r:s}$ --- i.e., which in phase space is most closely
located to the torus that lies symmetrically on the opposite side of
the resonance chain.  This contribution is particularly enhanced by a
small energy denominator [see Eq.~(\ref{Ediff})] and typically
dominates the sum in Eq.~(\ref{eq:split1}).

As one goes further in the semiclassical $\hbar \to 0$ limit, a
multi-resonance process is usually the dominant one. Neglecting
interference terms between different coupling pathways that connect
the ground state with a given excited mode $|n\rangle$ (which is
justified due to the fact that the amplitudes associated with those
coupling pathways are, in general, much different from each other in
size), we obtain from Eq.~(\ref{eq:multi_amp}) an expression of the form
\begin{equation}\label{eq:Deltaphimultires}
  \Delta \varphi_0 = \Delta \varphi_0^{(0)} + 
  \sum_k |\mathcal{A}_{0,kr}^{(r:s)}|^2 \Delta \varphi_{kr}^{(0)} + 
  \sum_k \sum_{k'} |\mathcal{A}_{0,kr}^{(r:s)}|^2 
  |\mathcal{A}_{kr,kr+k'r'}^{(r':s')}|^2 \Delta \varphi_{kr+k'r'}^{(0)} + \ldots
  \label{eq:split2}
\end{equation}
with the coupling amplitudes
\begin{eqnarray}\label{eq:Amplitudes1}
  \mathcal{A}_{0,kr}^{(r:s)} & = & \prod_{l=1}^k 
  \frac{\langle lr|\hat{H}_{\rm res}^{(r:s)}|(l-1)r\rangle}
  {E_0 - E_{lr} + l s \hbar \omega} \\\label{eq:Amplitudes2}
  \mathcal{A}_{kr,kr+k'r'}^{(r':s')} & = & \prod_{l'=1}^{k'}
  \frac{\langle kr + l'r' |\hat{H}_{\rm res}^{(r':s')}|kr + (l'-1)r'\rangle}
  {E_0 - E_{kr+l'r'} + (ks +l's') \hbar \omega} \\
  \mathcal{A}_{kr+k'r',kr+k'r'+k''r''}^{(r'':s'')} & = & \ldots \nonumber
\end{eqnarray}
for the  eigenphase splitting.

\subsection{Coupling with the chaotic sea}
\label{sec:CwCS}

Turning now to the mixed regular-chaotic case, the integrable
Hamiltonian $H_0(I)$
provides a good approximation only near the center
of the regular island under consideration, and invariant tori exist
only up to a maximum action variable $I_c$ corresponding to the
outermost boundary of the regular island in phase space.  Beyond this
outermost invariant torus, multiple overlapping resonances provide
various couplings and pathways such that unperturbed states in this
regime can be assumed to be strongly connected to each other.  Under
such circumstances, it is natural to divide the Hilbert space into two
parts, integrable and chaotic, associated respectively with the 
phase space regions within and outside the regular island. 

For each symmetry class $\pm$ of the problem, let us introduce an
effective Hamiltonian $\hat{H}^\pm_{\rm eff}$ modeling the tunneling
process.  Let us furthermore denote $\hat P_{\rm reg}$ and $\hat
P_{\rm ch}$ the (orthogonal) projectors onto the regular and chaotic 
Hilbert spaces.
The diagonal blocks $\hat{H}^\pm_{\rm reg} \equiv
\hat P_{\rm reg} \hat{H}^\pm_{\rm eff} \hat P_{\rm reg}$ and
$H^\pm_{\rm ch} \equiv \hat P_{\rm ch} \hat{H}^\pm_{\rm eff} \hat
P_{\rm ch}$ receive a natural interpretation: 
within $\hat{H}^\pm_{\rm reg}$, on the one hand,
the dynamics is exactly the same as in the nearly integrable regime above;
$\hat{H}^\pm_{\rm ch}$, on the other hand,
is best modeled in a statistical manner
by the introduction of random matrix ensembles.
The only remaining delicate point is thus to connect the two, namely
to model the off-diagonal block $\hat P_{\rm reg} \hat{H}^\pm_{\rm eff} \hat
P_{\rm ch}$.  We stress that there is not yet a real consensus on 
the best way how to do this, although various approaches give good 
quantitative accuracy.

To state more clearly the problem, let us consider a
regular state $|\bar{n}\rangle$
with quasi-energy $E_{\bar{n}}^0$ close to the regular-chaos boundary. 
(Note that ``close'' here means that no
resonance within the island can connect $|\bar{n}\rangle$ to a state
$|n'\rangle$ within the island with $n'>\bar{n}$.  This notion of
``closeness'' to the boundary is therefore $\hbar$-dependent.)  The
resonance assisted mechanism will connect any quasi-mode deep inside
the island to such a state at the edge of the island.  But to complete
the description of the chaos-assisted tunneling process it is
necessary to compute the variance $v^2_{\bar{n}}$ of the random matrix
elements $v_{\bar{n}i}$ between $| \bar{n} \rangle $ and the eigenstates $|
\psi^c_i\rangle $ of $\hat{H}^\pm_{\rm ch}$  (the variance is independent of
$i$ if $\hat{H}^\pm_{\rm ch}$ is modeled by the Gaussian orthogonal or
unitary ensemble).

One possible approach to compute this quantity is the fictitious 
integrable system approach that was proposed by
B\"acker et al.\ \cite{BaeO08PRL}.  This method relies on the fact
that, for the effective Hamiltonian $\hat{H}_{\rm eff}$, the ``direct''
transition rate from a
regular state $|n\rangle$ to the chaotic region is given using Fermi's
golden rule by
\begin{equation} \label{eq:FGR}
  \Gamma^{d}_{n\to {\rm chaos}} = \frac{2\pi}{\hbar} \frac{v_n^2}{
    \Delta_{\rm ch}}  \; , 
\end{equation}
(see e.g. section 5.2.2 of Ref.~\cite{BohTomUll93PR} for a
discussion in the context of random matrix theory) where
$\Delta_{\rm ch}$ denotes the 
mean spacing between eigenenergies within the chaotic block.
As a consequence, one obtains in first order in $\tau v_n^2/ \hbar 
\Delta_{\rm ch}$,
\begin{equation} \label{eq:SemiNum}
||\hat P_{\rm ch} 
\hat{U}
|n\rangle ||^2 = 
\tau \Gamma^{d}_{n\to {\rm chaos}} =  \frac{2\pi \tau}{\hbar} 
\frac{v_n^2}{ \Delta_{\rm ch}} \;
\end{equation}
(see also the contribution of B\"acker, Ketzmerick, and L\"ock in this book).
If one can explicitly construct 
a good integrable approximation $H_{\rm reg} \equiv H_0(p,q)$ 
of the time-dependent dynamics (see Sec.~\ref{sec:sec}), this
allows one, by quantum or semiclassical diagonalization, 
 to determine the unperturbed eigenstates $|n\rangle$ within the
 regular island, and to construct the projectors $\hat P_{\rm reg}$
 and $\hat P_{\rm ch}$ The ``direct'' regular-to-chaotic tunneling
 matrix elements of the $n$th quantized state within the island is
 then evaluated by a simple application of the quantum time evolution
 operator $\hat{U}$ over one period of the driving.

This approach can be qualified as ``semi-numerical'', as it requires
to numerically perform the quantum evolution for one period of the map
(although this can be done by analytical and semiclassical techniques
in some cases \cite{BaeO208PRL}, see also the contribution of B\"acker et 
al.\ in this book).
It strongly relies on the quality of the integrable approximation 
$\hat{H}_{\rm reg}$ of the Hamiltonian. 
If the latter was really diagonal (which, as a matter of principle, 
cannot be achieved by means of classical perturbation theory, due to the 
appearance of nonlinear resonances), Eq.~(\ref{eq:SemiNum}) would
represent an exact result (apart from the first-order approximation 
in $\tau v_n^2/ \hbar \Delta_{\rm ch}$ which should not be a limitation 
in the tunneling regime).
And indeed very good agreement between this prediction and
numerically computed tunneling rates was found for quantum maps that
were designed such as to yield a ``clean'' mixed regular-chaotic phase
space, containing a regular island and a chaotic region which both do
not exhibit appreciable substructures \cite{BaeO08PRL}, as well as for
the mushroom billiard \cite{BaeO208PRL}.  In more generic situations,
where nonlinear resonances are manifested within the regular island,
this approach yields reliable predictions for the direct tunneling of
regular states at the regular-chaos border, and its combination with 
the resonance assisted mechanism described in the previous sections 
leads to good quantitative predictions for the tunneling rates for the 
states deep in the regular island \cite{LoeO10PRL}.  We
shall illustrate this on the example of the kicked rotor system in
section \ref{sec:kr}.

\subsection{Integrable semiclassical models for the regular-to-chaotic 
coupling}

For now, however, we shall discuss other possible approaches of purely
semiclassical nature (i.e. not involving any numerical evolution nor 
diagonalization of the quantum system) to the calculation of the coupling
parameter $v_{\bar{n}}$ for a regular state $|\bar{n}\rangle$ at the edge of
the regular-chaos boundary.

One way to obtain an order of magnitude of
$v_{\bar{n}}^2$ is to consider the decay of the quasimode inside the regular
island. For this purpose, let us assume that an integrable
approximation $H_0(I)$, valid up to a maximum action $I_c$
corresponding to the chaos boundary, has been obtained.  Within the
Birkhoff-Gustavson normal-form coordinates $(P,Q)$ given by
Eqs.~(\ref{eq:BGp}) and (\ref{eq:BGq}), $H_0$ appears as a function of the
harmonic-oscillator Hamiltonian $(P^2+Q^2)/2$ and therefore has the
same eigenstates
\begin{eqnarray}
 \langle Q | \ {\bar{n}}\rangle & = & 
 \frac{1}{\sqrt{2^{\bar{n}} {\bar{n}}!}}
 \frac{1}{(\pi \hbar)^{1/4}} \exp\left( - Q^2/2\hbar \right) 
      {\cal H}_{\bar{n}}(Q/\hbar^{1/2}) \label{eq:oh_eigenstate} \\
      & \simeq & \frac{1}{2\sqrt{2\pi |P_{\bar{n}}(Q)|}} \exp\left(
      -\int_{Q_1}^Q|P_{\bar{n}}(Q)| dQ  \right) 
      \qquad \mbox{for}\quad Q>Q_n \; , \label{eq:oh_eigenstate_sc}
\end{eqnarray}
where ${\cal H}_{\bar{n}}$ are the Hermite polynomials and 
$P_{\bar{n}}(Q) = \sqrt{2I_{\bar{n}} - Q^2}$. 
The last equation corresponds to the semiclassical asymptotics in the 
forbidden region on the right-hand side of the turning point 
$Q_1 = \sqrt{2I_{\bar{n}}}$ with
$I_{\bar{n}} = \hbar({\bar{n}} +1/2)$.
In $Q$ representation, the regular island extends up to $Q_c
= \sqrt{2I_c}$, at which point de state $|{\bar{n}}\rangle$ has decayed to
$\psi_{\bar{n}}(Q_c) \simeq \frac{1}{2\sqrt{2\pi |P_{\bar{n}}(Q_c)|}}
\exp\left( -S(I_{\bar{n}},I_c) \right)$, where
\begin{equation} \label{eq:SInIc}
S(I_{\bar{n}},I_c) =\int_{Q_1}^{Q_c}|P_{\bar{n}}(Q)|   dQ
 =  \sqrt{I_c(I_c - I_{\bar{n}})} -  I_{n} \ln \left( (\sqrt{I_c - I_{\bar{n}}} +
   \sqrt{I_c} ) / \sqrt{I_{\bar{n}}} \right) \; .
\end{equation}
This expression is suspiciously simple (as it depends
only on $I_{\bar{n}}$ and $I_c$, and on no other property of the system), 
and should not be taken too seriously as it is.

Indeed, it should be borne in mind that $v_{\bar{n}}^2$ is related not so
much to the value of the wavefunction at the regular-chaos boundary
than to the transition rate $\Gamma^{d}_{n\to {\rm chaos}}$ through
Eq.~(\ref{eq:FGR}), which we can equal to the current flux $J_{\bar{n}}$
through this boundary for the regular state $|{\bar{n}}\rangle$.
Using $H_0(I)$ to compute this current leads to a zero result 
and it is therefore
mandatory to use a better approximation of the non-integrable
Hamiltonian $H$ to obtain a meaningful answer.  What complicates
the evaluation of the transition rate from the edge of the
regular to the chaotic domain is therefore that one needs to find an 
approximation describing both the regular and chaotic dynamics 
--- unless one actually uses
there the exact quantum dynamics as was done in Ref.~\cite{BaeO08PRL}.
Since the regular-chaos border is typically the place where
approximation schemes tend to be difficult to control, this will rely
on some assumption to be made for the chaotic regions, two possible 
choices of which we shall describe now.

\begin{figure}[ht]
  \begin{center}
    \includegraphics*[width=6cm]{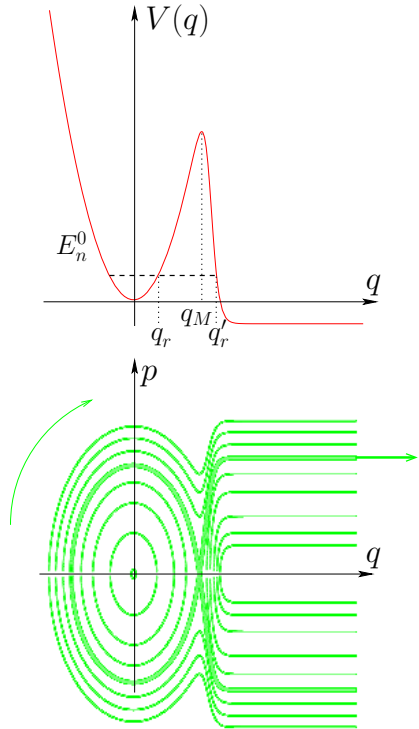}
  \end{center}
  \caption{Modeling of the direct coupling between the edge of the
    regular region to the chaotic one by a potential barrier
    separating a potential well from an ``open'' region.  (a) Sketch
    of the potential.  (b) Corresponding phase space portrait.
    \label{fig:vescape}
  }
\end{figure}

One scenario that has been considered amounts in some way to
model the regular to chaos transition in the way depicted in
Fig.~\ref{fig:vescape}a: a kinetic-plus-potential 
Hamiltonian $p^2/2m+V(q)$ where the
island itself correspond to a potential well and the edge of the
regular region to the place where the potential decreases
abruptly.  The picture one has in mind in that case is that
escaping from the edge of the regular island to the chaotic sea is
akin to the standard textbook barrier tunneling
\cite{Merzbacher70a}. 
Using Langer's connection formula \cite{Langer} within that model, 
the semiclassical wavefunction for the quasi-mode ${\bar{n}}$ inside 
the potential well can be extended under the potential 
barrier and, beyond this, into the region where motion at energy 
$E^0_{\bar{n}}$ is again classically authorized.  
In the classically allowed region outside the well ($q>q'_r$) 
the semiclassical wavefunction can be written as
\begin{equation} \label{eq:semibeyond}
\psi_{\bar{n}}(q) \simeq \sqrt{\frac{ \Omega_{\bar{n}}}{2\pi p_{\bar{n}}(q)}}
  \exp\left(\frac{i}{\hbar} \int_{q'_r}^q  p_{\bar{n}}(q) dq + i \pi/4 \right)
   \exp \left( - \frac{S_{\rm t}}{\hbar} \right)\; ,
\end{equation}
with $\Omega_{\bar{n}}$ the angular frequency of the torus
$E_{\bar{n}}^0$, $p_{\bar{n}}(q) = \sqrt{2m\big[E_{\bar{n}}^0 - V(q)\big]}$, and  
\begin{equation} \label{eq:tunnelingS} 
S_{\rm t} = \int_{q_r}^{q'_r} |p_{\bar{n}}(q)| dq
\end{equation}
the tunneling action\footnote{Strictly speaking, 
  outgoing (Siegert) boundary conditions \cite{Sie39PR}
  need to be employed in Eq.~(\ref{eq:semibeyond}) in order to properly
  describe the decay process from the well. Those outgoing boundary conditions
  involve, in addition, an exponential increase of the wavefunction's amplitude
  with increasing distance from the well, which is not taken into account in
  Eq.~(\ref{eq:semibeyond}) assuming that the tunneling rate from the well is
  comparatively weak.}.
The current of probability leaving the well is then given by
\begin{equation} \label{eq:In} 
J_{\bar{n}} \equiv \frac{\hbar}{m}\, {\rm Im} [\psi_{\bar{n}}^*
  \nabla \psi^{}_{\bar{n}}] = \frac{\Omega_{\bar{n}}}{2\pi}\exp 
\left( - 2 \frac{S_{\rm t}}{\hbar} \right) \; ,
\end{equation}
from which $v^2_{\bar{n}}$ is obtained through Eq.~(\ref{eq:FGR})
(identifying $J_{\bar{n}}$ with $\Gamma^{d}_{\bar{n} \to {\rm chaos}}$).

Although Eq.~(\ref{eq:In}) is derived here for the particular case of
a kinetic-plus-potential Hamiltonian, it applies more generally 
(up maybe to factors of order one) to any system with a phase space
portrait that is similar to the one of Fig.~\ref{fig:vescape}b, 
where tori inside the island can be analytically continued in the complex 
plane to a manifold escaping to infinity.  In that case, Eq.~(\ref{eq:In})
can be applied provided the tunneling action $S_t$ is taken as the
imaginary part of the action integral on a path joining the interior
to the exterior of the island on this analytical continuation.  As a
last approximation, one may assume that the transition to the ``open''
part is extremely sharp once the separatrix is crossed.  In the model
of Fig.~\ref{fig:vescape}a, this amounts to assume a very rapid 
decrease of the potential, in which case one may replace ${q_r}$ 
by ${q_M}$ in the tunneling action Eq.~(\ref{eq:tunnelingS}).
In an actual calculation of the direct tunneling rate for a regular 
state at the edge of the chaos boundary (which we can reliably describe 
only coming from the interior of the island) this amounts to consider 
in the same way that the action $S(I_{\bar{n}},I_c)$ [Eq.~(\ref{eq:SInIc})]
provides a good approximation to $S_t$.  Under this hypothesis, one
obtains for the coupling to the chaotic Hilbert space the prediction
\begin{equation} \label{eq:Vn} 
v^2_{\bar{n}} = \frac{\Delta_{\rm ch}\hbar \Omega_{\bar{n}}}{4\pi^2}
\exp \left( - 2 \frac{S(I_{\bar{n}},I_c)}{\hbar} \right) \; 
\end{equation}
(see \cite{LeDeunff/Mouchet10a} where this computation was
proposed with the slightly different language of complex time
trajectories).

This ``potential-barrier'' picture of the direct tunneling is in
essence what is behind the approach of Podolskiy and Narimanov
\cite{PodNar03PRL} (though their treatment of the problem is a bit more 
sophisticated). Its main virtue is that, beyond the
quantized action $I_{\bar{n}} = \hbar({\bar{n}}+1/2)$, 
Eq.~(\ref{eq:Vn}) relies only on simple characteristics of the
integrable regions : its area $2\pi I_c$ and the frequency
$\Omega_{\bar{n}}$ of the torus $I_{\bar{n}}$.  One needs to keep
in mind, however, that Eq.~(\ref{eq:Vn}) implicitly assumes 
that the direct tunneling mechanism corresponds in some sense to the
phase portrait of Fig.~\ref{fig:vescape}b, which in general cannot be
justified within any controlled approximation scheme.

\subsection{Regular-to-chaotic coupling via a nonlinear resonance}
\label{sec:RCviaNLR}

Another possible, and presumably more realistic, approach to
evaluate the direct coupling parameter $v_{\bar{n}}$ can be
obtained assuming that the effective model (\ref{heff}) describes the
vicinity of a resonance not only inside the regular region, but also
within the chaotic sea in the near vicinity of the island. In this
perspective, the model one has in mind for the chaotic region
follows the spirit of Chirikov's overlapping criterion for the
transition to chaos \cite{Chirikov,LicLie}: 
resonances still provide the couplings between quasi-modes, but 
in the chaotic region these couplings become strong enough to 
completely mix the states.  Near the regular-chaos edge, the 
transition between modes inside and outside the regular region 
is still dominated by one or several $r$:$s$ resonances which 
might be within or possibly outside the regular island.

As an illustration, let us consider the simple case where a single
nonlinear $r$:$s$ resonance is responsible for all couplings, both
within the island and from the island's edge to the chaotic region.
Keeping in mind the discussion in Section \ref{sec:pend} and in
Ref.~\cite{BroSchUll02AP}, we assume here that the couplings induced
by the $r$:$s$ resonance are dominantly described by the lowest
non-vanishing Fourier component $V_1$ of the perturbation, 
i.e.\ by the matrix elements
$V_{r:s}^{(n+r)} \equiv \langle n + r|\hat{H}_{\rm eff}|n\rangle$, 
and set the phase $\phi_1$ to zero without loss of generality.

The structure of the effective Hamiltonian that describes the coupling
of the ground state $E_0$ to the chaotic sea is, in that case, given by
\begin{equation}
    \newlength{\Vrs}
    \settowidth{\Vrs}{$V_{r:s}$}
    \newlength{\csize}
    \setlength{\csize}{6.5em}
    H^\pm_{\rm eff} = 
    \left(
      \begin{array}{ccc@{\hspace*{1cm}}l}
        \tilde E_0    & V_{r:s}^{(r)} & \phantom{\ddots}     &  \\
        V_{r:s}^{(r)} &  \ddots       &  {\ddots}    &  \\
        & \ddots      & \tilde{E}_{k_cr}
                      & \hspace*{-0.3cm} V_{r:s}^{[(k_c+1) r]} \\
        [2mm]         &               & 
           \parbox[c][\csize][t]{\Vrs}{$V_{r:s}^{[(k_c+1) r]}$} &
        \framebox{\parbox[c][\csize][c]{\csize}{\centering\large chaos$^\pm$}}
       \\
      \end{array}
    \right) \; , \label{elabet}
\end{equation}
where $\tilde{E}_{kr} \equiv E_{kr} - \Omega_{r:s} I_{kr}$ are the
eigenenergies of the unperturbed Hamiltonian $\mathcal{H}_0$ in the
co-rotating frame, and the chaotic part (the square in the lower right 
corner) consists of a full sub-block with equally strong couplings 
between all basis states with actions beyond the outermost invariant 
torus of the islands. In this example the last state within the island 
connected to the ground state is the quasi-mode $\bar{n} = k_cr$
and $v^2_{\bar{n}} = (V_{r:s}^{[(k_c+1) r]})^2 / N_{\rm ch}$, 
with $N_{\rm ch}$ the number of
states with a given parity in the chaotic Hilbert space. 
In a more general situation, many resonances may, as in
Eq.~(\ref{eq:multi_amp}), be involved in connecting the ground stated
to some $|\bar{n}\rangle$ at the edge of the island, but we would still
write the variance $v^2_{\bar{n}}$ of the matrix elements providing the
last coupling to the chaotic region as the ratio 
$(V_{r:s}^{[(k_c+1) r]})^2/N_{\rm ch}$ for some $r$:$s$ resonance 
near the regular-chaos edge.

We stress here that it is necessary in this approach to have an explicit 
access to the number of states $N_{\rm ch}$ in the chaotic
Hilbert space.  This is obtained quite trivially for quantum
maps, such as the kicked rotor we shall consider in
section~\ref{sec:kr}, when $\hbar = 2\pi/N$ with integer $N$.  
In that case $N$ is the total number of states in the
full Hilbert space and~$N_{\rm ch}/N$ represents
the relative area of the chaotic region in phase space.
For a two dimensional conservative Hamiltonian, on the other
hand, $N$ and thus $N_{\rm ch}$ are related to the Thouless energy 
(see e.g.\ the section 2.1.4 of \cite{UllmoRPP2008}), 
but this provides only an energy scale rather than a precise number, 
and a more detailed discussion is required to be quantitative.  
For the sake of clarity, we shall in the following limit our discussion 
to the simpler case that $N$ is known.

\subsection{Theory of chaos-assisted tunneling}

\label{sec:tcat}

Let us consider now the effect of the chaotic block on the tunneling
process. Eliminating intermediate states within the regular island
leads for the effective Hamiltonian a matrix of the form
\begin{equation}
  H_{\rm eff}^\pm = \left( 
    \begin{array}{ccccc}
      E_0 & V_{\rm eff} & 0 & \cdots & 0 \\
      V_{\rm eff} & H_{11}^\pm & \cdots & \cdots & H_{1N_{\rm ch}}^\pm \\
      0 & \vdots & & & \vdots \\
      \vdots & \vdots & & & \vdots \\
      0 & H_{N_{\rm ch}1}^\pm & \cdots & \cdots & H_{N_{\rm ch}N_{\rm ch}}^\pm
    \end{array}
  \right) \, .
  \label{eq:model}
\end{equation}
for each symmetry class.  In the simplest case Eq.~(\ref{elabet})
where a single $r$:$s$ resonance needs to be considered, the effective
coupling matrix element between the ground state and the chaos block
$(H_{ij}^\pm)$ is given by
\begin{equation}\label{eq:Veff}
  V_{\rm eff} = V_{r:s}^{[(k_c+1) r]} \prod_{k=1}^{k_c}\frac{V_{r:s}^{(kr)}} 
  {E_0 - E_{kr} + k s \hbar \omega} \label{eq:veff}
\end{equation}
where $E_n$ are the unperturbed energies (\ref{eq:en}) of $H_{\rm eff}$ 
and $|k_c r\rangle$ represents the highest unperturbed state
that is connected by the $r$:$s$ resonance to the ground state and
located within the island (i.e., $I_{k_c r} < I_c < I_{(k_c+1) r}$).
More generally, $V_{\rm eff}$ can be expressed in terms of the couplings 
associated with the various resonances that contribute to the transitions
within the island and at the regular-chaos edge.
The form of this expression (\ref{eq:veff}) already provides an
explanation for the appearance of plateau-like structures in the
tunneling rates.
Indeed, decreasing $\hbar$ leads to discontinous increments of the 
maximal number $k_c$ of couplings through Eq.~(\ref{eq:kc}) and hence
to step-like reductions of the effective matrix element $V_{\rm eff}$,
while in between such steps $V_{\rm eff}$ varies smoothly through the action
dependence of the coupling matrix elements $V_{r:s}^{(kr)}$, provided
accidential near-degeneracies in the energy denominators do not occur.

In the simplest possible approximation, which follows the lines of
Refs.~\cite{TomUll94PRE,LeyUll96JPA}, we neglect the effect of partial
barriers in the chaotic part of the phase space \cite{BohTomUll93PR} and
assume that the chaos block $(H_{ij}^\pm)$ is adequately modeled by a random
Hermitian matrix from the Gaussian orthogonal ensemble (GOE).
After a pre-diagonalization of $(H_{ij}^\pm)$, yielding the eigenstates 
$\phi_j^\pm$ and eigenenergies $\mathcal{E}_j^\pm$, we can perturbatively 
express the shifts of the symmetric and antisymmetric ground state energies by
\begin{equation}
  E_0^\pm = E_0 + \sum_{j=1}^{N_{\rm ch}} \frac{|v^j_{{\rm eff}\pm}|^2}{E_0 - 
    \mathcal{E}_j^\pm} \; ,
\end{equation}
with $v^j_{{\rm eff}\pm} \equiv V_{\rm eff} \langle kr|\phi_j^\pm \rangle$
Performing the random matrix average for the eigenvectors, we obtain that 
$ \langle \langle |\langle kr|\phi_j^\pm\rangle|^2 \rangle \rangle 
\simeq 1/N_{\rm ch}$ for all $j=1\ldots N_{\rm ch}$, which simply
expresses the fact that none of the basis states is distinguished
within the chaotic block $(H_{ij}^{\pm})$.  As a consequence, the variance
of the $v^j_{{\rm eff}\pm}$'s is independent of $j$ and equal to
$v^2_{\rm eff} = V^2_{\rm eff}/N_{\rm ch}$.

As was shown in Ref.~\cite{LeyUll96JPA}, the random matrix average over the
eigenvalues $\mathcal{E}_j^\pm$ gives rise to a Cauchy distribution for the 
shifts of the ground state energies, and consequently also for the splittings
\begin{equation}
  \Delta E_0 = |E_0^+ - E_0^-|
\end{equation}
between the symmetric and the antisymmetric ground state energy.
For the latter, we specifically obtain the probability distribution
\begin{equation}
  P(\Delta E_0) = \frac{2}{\pi} \frac{\overline{\Delta E_0}}
  {(\Delta E_0)^2 + (\overline{\Delta E_0})^2} \label{cauchy}
\end{equation}
with
\begin{equation}
  \overline{\Delta E_0} = \frac{2\pi v_{\rm eff}^2}{\Delta_{\rm ch}}
  \label{split}  
\end{equation}
where $\Delta_{\rm ch}$ denotes the mean level spacing in the chaos at
energy $E_0$.  This distribution is, strictly speaking, valid only for
$\Delta E_0 \ll v_{\rm eff}$ and exhibits a cutoff at $\Delta E_0 \sim
2 v_{\rm eff}$, which ensures that the statistical expectation value
$\langle\Delta E_0\rangle = \int_0^\infty x P(x) dx$ does not diverge.

Since tunneling rates and their parametric variations are typically studied 
on a logarithmic scale [i.e., $\log (\Delta E_0)$ rather than 
$\Delta E_0$ is plotted vs.\ $1/ \hbar$], 
the relevant quantity to be calculated from Eq.~(\ref{cauchy}) 
and compared to quantum data is not the mean value $\langle\Delta E_0\rangle$, 
but rather the average of the logarithm of $\Delta E_0$.
We therefore define our ``average'' level splitting 
$\langle\Delta E_0\rangle_g$ as the {\em geometric} mean of $\Delta E_0$, i.e.
\begin{equation}\label{eq:DeltaE_av}
  \langle\Delta E_0\rangle_g \equiv \exp \left[ \left\langle
      \ln(\Delta E_0) \right\rangle \right] 
\end{equation}
and obtain as result the scale defined in Eq.~(\ref{split}),
\begin{equation} \label{eq:geommean}
  \langle\Delta E_0\rangle_g = \overline{\Delta E_0} \, .
\end{equation}

This expression further simplifies for our specific case of periodically
driven systems, where the time evolution operator $\hat{U}$ is modeled by the
dynamics under the effective Hamiltonian (\ref{eq:model}) 
over one period $\tau$.
In this case, the chaotic eigenphases $\mathcal{E}_j^\pm \tau / \hbar$
are uniformly distributed in the interval $[0,2\pi[$. We therefore obtain
\begin{equation}
  \Delta_{\rm ch} = \frac{2 \pi \hbar}{N_{\rm ch} \tau}
\end{equation}
for the mean level spacing near $E_0$.
This yields
\begin{equation}
  \langle\Delta \varphi_0\rangle_g \equiv \frac{\tau}{\hbar} 
  \langle\Delta E_0\rangle_g = \left( \frac{\tau V_{\rm eff}}{\hbar} \right)^2
  \label{eq:splitg}
\end{equation}
for the geometric mean of the ground state's eigenphase splitting.
Note that this final result does not depend on the number $N_{\rm ch}$ of
chaotic states within the sub-block $(H_{ij}^\pm)$; as long as this number
is sufficiently large to justify the validity of the
Cauchy distribution (\ref{cauchy}) (see Ref.~\cite{LeyUll96JPA}), the
geometric mean of the eigenphase splitting is essentially given by the
square of the effective coupling $V_{\rm eff}$ 
from the ground state to the chaos.

The distribution (\ref{cauchy}) also permits the calculation of the
logarithmic variance of the eigenphase splitting: we obtain
\begin{equation}
  \left\langle \left[ \ln(\Delta \varphi_0) - \langle \ln(\Delta
      \varphi_0) \rangle \right]^2 \right\rangle  
  = \frac{\pi^2}{4} \, . \label{eq:var}
\end{equation}
This universal result predicts that the actual splittings may be enhanced or
reduced compared to $\langle\Delta \varphi_0\rangle_g$ 
by factors of the order of $\exp(\pi/2) \simeq 4.8$,
independently of the values of $\hbar$ and external parameters.
Indeed, as was discussed in Ref.~\cite{SchEltUll06},
short-range fluctuations of the splittings, 
arising at small variations of $\hbar$, are well characterized
by the standard deviation that is associated with Eq.~(\ref{eq:var}).

It is interesting to note that the expression (\ref{eq:splitg}) for
the (geometric) mean level spacing is quantitatively identical with the 
expression (\ref{eq:SemiNum}) for the mean escape rate from the regular
island to the chaotic sea derived in Ref.~\cite{BaeO08PRL} using 
Fermi's golden rule.
This seems surprising as two different nonclassical processes, namely 
Rabi oscillations between equivalent islands and the decay from an 
island within an open system, underly these expressions.
In one-dimensional single-barrier tunneling problems, these two processes
would indeed give rise to substantially different rates;
in Eq.~(\ref{eq:1dsplit}), to be more precise, the imaginary action integral
in the exponent would have to be multiplied by two in order to obtain the
corresponding expression for the decay rate (and the overall prefactor in 
front of the exponential function should be divided by two, which is not 
important here).
The situation is a bit different, however, in our case of dynamical tunneling
in mixed regular-chaotic systems.
In such systems, level splittings between two equivalent regular islands 
involve \emph{two} identical dynamical tunneling processes between the islands 
and the chaotic sea (namely one process for each island), while the decay
into the chaotic sea involves only \emph{one} such process, 
with, however, the square of the corresponding (exponentially suppressed) 
coupling coefficient.
This explains from our point of view the equivalence of the expressions
(\ref{eq:splitg}) and (\ref{eq:SemiNum}).

We finally remark that the generalization of the expression for the mean
splittings to multi-resonance processes is straightforward and amounts to
replacing the product of admixtures in Eq.~(\ref{eq:veff}) by a product
involving several resonances subsequently, in close analogy with
Eq.~(\ref{eq:Deltaphimultires}).
In fact, the multi-resonance expression (\ref{eq:Deltaphimultires})
can be directly used in this context replacing the ``direct'' splittings
$\Delta \varphi_{n}^{(0)}$ by $(V_{r_f:s_f}^{[(k_c+1) r_f]} \tau / \hbar)^2$
where the $r_f$:$s_f$ resonance is the one that induces the final coupling 
step to the chaotic sea (provided $I_n < I_c < I_{n + r_f}$ holds for the
corresponding action variables; otherwise we would set 
$\Delta \varphi_{n}^{(0)} = 0$).
This expression represents the basic formula that is used in the 
semiclassical calculations of the splittings in the kicked rotor model, 
to be discussed below.

\subsection{The role of partial barriers in the chaotic domain}

In the previous section, we assumed a perfectly homogeneous structure
of the Hamiltonian outside the outermost invariant torus, which
allowed us to make a simple random-matrix ansatz for the chaotic block.
This assumption hardly ever corresponds to reality.  As was shown in
Refs.~\cite{BohTomUll90PRL,BohTomUll93PR,TomUll94PRE} for the quartic
oscillator, the chaotic part of the phase space is, in general,
divided into several subregions which are weakly connected to each
other through partial transport barriers for the classical flux (see,
e.g., Fig.\ 8 in Ref.~\cite{TomUll94PRE})
This substructure of the
chaotic phase space (which is generally not visible in a Poincar\'e
surface of section) is particularly pronounced in the immediate
vicinity of a regular island, where a dense hierarchical sequence of
partial barriers formed by broken invariant tori and island chains is
accumulating \cite{MacMeiPer84PRL,MacMeiPer84PhD,MeiOtt85PRL}.

In the corresponding quantum system, such partial barriers may play the role
of ``true'' tunneling barriers in the same spirit as invariant classical tori.
This will be the case if the phase space area $\Delta W$ that is exchanged 
across such a partial barrier within one classical iteration is much smaller
than Planck's constant $2 \pi \hbar$ \cite{MaiHel00PRE}, while in the 
opposite limit $\Delta W \gg 2 \pi \hbar$ the classical partial barrier 
appears completely transparent in the quantum 
system\footnote{More precisely, the authors of Ref.~\cite{MaiHel00PRE} 
  claim that $\Delta W$ has to be compared with $\pi \hbar$ in order 
  to find out whether or not a given partial barrier is transparent 
  in the quantum system.}.
Consequently, the ``sticky'' hierarchical region around a regular island
acts, for not extremely small values of $\hbar$, as a dynamical tunneling
area and thereby extends the effective ``quantum'' size of the island in
phase space.
As a matter of fact, this leads to the formation of localized 
states (also called ``beach'' states in the literature \cite{DorFri95PRL})
which are supported by this sticky phase space region in the surrounding 
of the regular island \cite{KetO00PRL}
(see Fig.~\ref{fig:kickedrotor_husimi}b).

\begin{figure}[ht]
  \begin{center}
    \includegraphics*[width=\textwidth]{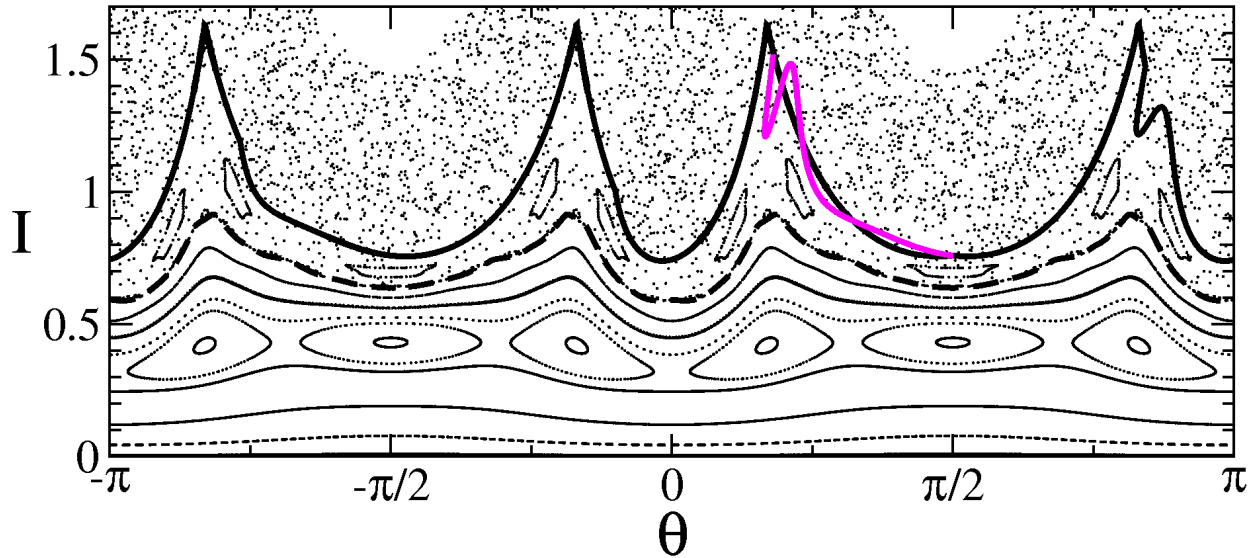}
  \end{center}
  \caption{
    Classical phase space of the kicked rotor at $K = 3.4$
    in approximate action-angle variables $(I,\theta)$.
    The thick solid line shows the location of the effective quantum 
    boundary of the central island for the values of Planck's 
    constant that are considered in Section \ref{sec:kr}.
    This effective boundary is constructed from segments of 
    the stable and unstable manifolds that emerge from the 
    hyperbolic periodic points of the 4:1 resonance 
    at $\theta \simeq \pi/6$ and $5 \pi/6$, respectively.
    Those segments were computed until the symmetry axis at 
    $\theta = \pi/2$ and then iterated three times under the
    kicked rotor map, yielding the thick solid line.
    A further iteration of this boundary 
    maps it onto itself, except for the piece between 
    $\theta \simeq 0.15 \pi$ and $\theta = 0.5 \pi$ 
    which is replaced by the lighter curve.
    The phase space area that is enclosed between the original (dark) and 
    the iterated (light) boundary defines the classical flux that is 
    exchanged across this boundary within one iteration of the map.
    The dashed line shows, in comparison, the actual classical
    chaos border defined by the outermost invariant torus of the island.
    \label{fg:border}
  }
\end{figure}

An immediate consequence of the presence of such partial barriers 
for resonance-assisted tunneling is the fact that the critical action 
variable $I_c$ defining the number $k_c$ of resonance-assisted steps
within the island according to Eq.~(\ref{eq:veff}) should not be determined
from the outermost invariant torus of the island, but rather from the
outermost partial barrier that acts like an invariant torus in the quantum
system.
We find that this outermost quantum barrier is, for not extremely small
values of $\hbar$, generally formed by the stable and unstable manifolds
that emerge from the hyperbolic periodic points associated with a low-order
nonlinear $r$:$s$ resonance.
These manifolds are constructed until their first intersection points 
in between two adjacent periodic points, and iterated $r-1$ times 
(or $r/2-1$ times in the case of period-doubling of the island chain 
due to discrete symmetries), such as to form a closed artificial 
boundary around the island in phase space\footnote{This 
  construction is also made in order to obtain the phase space areas
  $S_{r:s}^\pm$ that are enclosed by the outer and inner separatrix structures
  of an $r$:$s$ resonance, and that are needed in order to compute the mean
  action variable $I_{r:s}$ and the coupling strength $V_{r:s}$ of the
  resonance according to Eqs.~(\ref{eq:sep}) and (\ref{eq:trm}).} 
As shown in Fig.~\ref{fg:border}, one further iteration maps
this boundary onto itself, except for a small piece that develops a
loop-like deformation.  The phase space area enclosed between the
original and the iterated boundary precisely defines the classical
flux $\Delta W$ exchanged across this boundary within one iteration of
the map \cite{MacMeiPer84PRL,MacMeiPer84PhD}.

The example in Fig.~\ref{fg:border} shows a boundary that arises from
the inner stable and unstable manifolds (i.e.\ the ones that would, in
a near-integrable system, form the inner separatrix structure)
emerging from the unstable periodic points of a 4:1 resonance (which
otherwise is not visible in the Poincar\'e section) in the kicked
rotor system.  Judging from the size of the flux area $\Delta W$, this
boundary should represent the relevant quantum chaos border for the
tunneling processes that are discussed in the following section.  We
clearly see that it encloses a non-negligible part of the chaotic
classical phase space, which includes a prominent 10:3 resonance that,
consequently, needs to be taken into account for the coupling process
between the regular island and the chaotic sea.  Thereby, we
naturally arrive at \emph{multi-step} coupling processes across a
sequence of several resonances, which would have to be computed for a
reliable prediction of the tunneling rates in the semiclassical
regime.

\section{Application to the kicked rotor}

\label{sec:kr}

\subsection{Tunneling in the kicked rotor}

To demonstrate the validity of our approach, we apply it to the 
``kicked rotor'' model, which is described by the Hamiltonian
\begin{equation}
  H(p,q,t) = p^2/2 - K \sum_{n=-\infty}^\infty \delta( t - n ) \cos q \, .
  \label{eq:kr}
\end{equation}
The classical dynamics of this system is described by the ``standard map''
$(p,q) \mapsto (p',q')$ with
\begin{eqnarray}
  p' & = & p - K \sin q \label{eq:map1} \\
  q' & = & q + p' \, , \label{eq:map2}
\end{eqnarray}
which generates the stroboscopic Poincar{\'e} section at times
immediately before the kick.  The phase space of the kicked rotor is
$2\pi$ periodic in position $q$ and momentum $p$, and exhibits, for
not too large perturbation strengths $K < 4$, a region of bounded
regular motion centered around $(p,q) = (0,0)$.

The quantum dynamics of the kicked rotor is described by the
associated time evolution operator
\begin{equation}
  \hat{U} = \exp \left( - \frac{i}{\hbar} \frac{\hat{p}^2}{2} \right) 
  \exp \left( - \frac{i}{\hbar} K \cos \hat{q} \right)
\end{equation}
which contains two unitary operators that describe the effect of the kick 
and the propagation in between two kicks, respectively 
($\hat{p}$ and $\hat{q}$ denote the position and momentum operators).

Because of the classical periodicity in both $p$ and $q$, we can consider 
tunneling between the main regular island centered around $(p,q)=(0,0)$ 
and its counterparts that are shifted by integer multiples of $2 \pi$ 
along the momentum axis or along the $q$-axis.
To mimic a double-well configuration, we will restrict the boundary 
conditions for the eigenstates of~$\hat{U}$ and consider tunneling between 
two islands centered around (i) $(p,q)=(0,0)$ and $(2\pi,0)$ or around
(ii) $(p,q)=(0,0)$ and $(0,2\pi)$.
The effective parity that allows to discriminate the eigenphases
$\varphi^\pm_n$ of $\hat{U}$ manifests 
as~$\tilde{\psi}^\pm_n(p+2\pi)=\pm\tilde{\psi}^\pm_n(p)$ for the 
corresponding eigenstates in momentum representation in case (i) and
as~$\psi^\pm_n(q+2\pi)=\pm\psi^\pm_n(q)$ for the 
eigenstates in  position representation in case (ii). 
In both cases, tunneling will be characterized by the splitting
\begin{equation}
  \Delta \varphi_n = |\varphi_n^{+} - \varphi_n^{-}| \;.\label{eq:split}
\end{equation}
Numerically, it can be convenient to deal a finite Hilbert space of
(even) size~$N$, and this can be obtained provided the two phase space
translation operators $\hat{T}_1 = \exp( 2 \pi i \hat{p} / \hbar )$
and $\hat{T}_2 = \exp( -2 \pi i \hat{q} / \hbar )$ commute, 
which is the case if we choose $\hbar = 2 \pi / N$
\cite{IzrShe79SPD,FisGuaReb02PRL}.

\subsection{Numerical computation of the eigenphase splittings}

\begin{figure}[!ht]
  \begin{center}
    \includegraphics*[width=8.5cm]{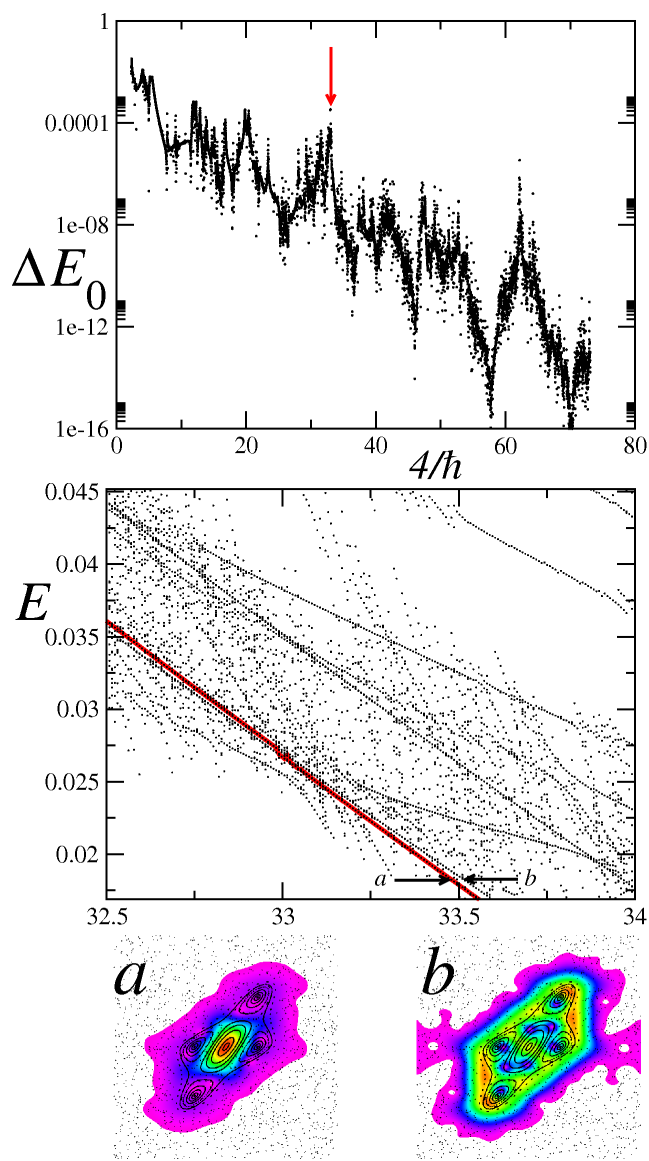}
  \end{center}
  \caption{ The upper panel shows the quasi-energy splittings
    in~$\Delta E_0=\hbar\varphi_0$ in the kicked rotor model for
    $K = 2.28$, corresponding to a $4:1$ classical resonance. Here we
    are concerned with a tunneling in~$q$.  The middle panel shows the
    quasi-energy spectrum~$E_n^\pm=\hbar\varphi_n^\pm$
    of~$\hat{U}$ where only states with a significant overlap with a
    coherent state localized around~$(p,q)=(0,0)$ have been retained.
    The horizontal arrow a) marks the central state doublet ($n=0$,
    not resolved at that scale) and the arrow b) indicates the third
    excited state localized in the island ($n=4$). Their Husimi
    distribution superimposed with the Poincar\'e surface of section
    are shown in the lower panel in order to illustrate the clear
    correspondence between the classical and the quantum resonance.
    As $1/\hbar$ increases, the crossing of the doublet by the
    resonant state provokes the large and wide spike indicated by the
    vertical arrow in the upper panel.
    \label{fig:kickedrotor_husimi}
  }
\end{figure}

\begin{figure}[!ht]
  \begin{center}
    \includegraphics*[width=10cm]{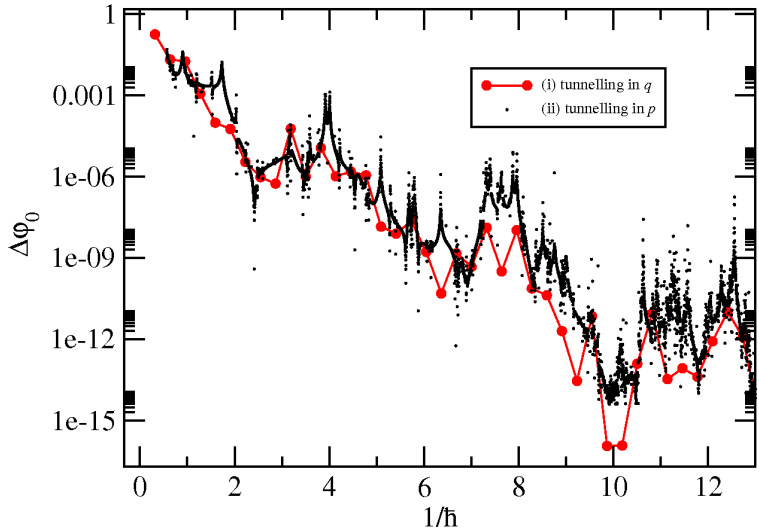}
  \end{center}
  \caption{\label{fig:comparaison_i_ii}Comparison between the
    splitting~$\Delta\varphi_0$ for the kicked rotor at $K=2$ for 
    (i) the case of tunneling between two islands centered at 
    $(p,q)=(0,0)$ and at $(p,q)=(2\pi,0)$, and 
    (ii) the case of tunneling between two islands
    centered at $(p,q)=(0,0)$ and at $(p,q)=(0,2\pi)$.  
    In this latter case, we restricted ourselves to even integer
    values of $N = 2 \pi / \hbar$ for which the splittings can be
    computed  by diagonalizing finite $N\times N$ matrices
    beyond the double precision.
}
\end{figure}

Figure~\ref{fig:kickedrotor_husimi} shows the eigenphase splittings 
$\Delta \varphi_0$ [see Eq.~(\ref{eq:split})] in case (ii) 
(tunneling in position) for the local ``ground state'' ($n=0$) in the
central island of the kicked rotor, i.e.\ for the state that is most 
strongly localized around the center of the island, at $K = 2.28$. 
While on average   these splittings decrease exponentially with $1/\hbar$, 
significant fluctuations arise on top of that exponential decrease.  
In particular, large spikes are visible.
As illustrated in Fig.~\ref{fig:kickedrotor_husimi}b, they can be 
related to the crossing of ``excited states'' within the island, 
which are coupled to the ground state by a classical resonance.
Fig.~\ref{fig:kickedrotor_husimi}c shows the Husimi distribution
of the relevant states involved, demonstrating that the coupling
process is most effective when the states are symmetrically
located on each side of the classical resonance.  To illustrate
that the influence of the resonances inside the regular islands is
actually independent of the details of the chaotic regime (a major
feature of the resonance-assisted and chaos-assisted tunneling
schemes) we furthermore plot in Fig.~\ref{fig:comparaison_i_ii} 
a comparison between the splittings for the cases (i) and (ii) of
tunneling in $p$ and $q$ direction, respectively.
The (rough) matching between the dominant spikes of fluctuations 
in both cases confirms that tunneling outside one island is mainly 
isotropic in phase space.

From now on, we will consider case (i) only (tunneling in momentum) 
with $N=2 \pi / \hbar$ being an integer, corresponding to the number 
of Planck cells that fit into one Bloch cell.  
Figures \ref{fg:kr1} and \ref{fg:kr2} show the eigenphase splittings 
$\Delta \varphi_0$ of the island's ground state for $K=2.6$, $2.8$, $3.0$ 
(Fig.~\ref{fg:kr1}) as well as for $K=3.2$, $3.4$, $3.6$ (Fig.~\ref{fg:kr2}).
As in Refs.~\cite{EltSch05PRL,SchEltUll06}, these splittings
were calculated with a diagonalization routine for complex matrices
that is based on the GMP multiple precision library \cite{gmp}, in
order to obtain accurate eigenvalue differences below the ordinary
machine precision limit.

\subsection{Semiclassical calculations}

The role played by the nonlinear resonances in the tunneling mechanism 
is made explicit by comparing these numerically calculated splittings 
with semiclassical predictions based on the most relevant resonances 
that are encountered in phase space.
In practice, we took those $r$:$s$ resonances into account that exhibit 
the smallest possible values of $r$ and $s$ for the winding numbers 
$s/r$ under consideration. In all of the considered cases, the 
``quantum boundary'' of the regular island, which determines the value 
of~$k_c$ through Eq.~(\ref{eq:kc}), was defined by the partial barrier 
that results from the intersections of the inner stable and unstable 
manifolds associated with the hyperbolic periodic points of the 
4:1 resonance (see also Fig.~\ref{fg:border}).  
While this partial barrier lies rather close to the classical chaos border 
of the island for $K = 2.6$ (Fig.~\ref{fg:kr1}), it encloses an appreciable 
part of the chaotic sea for $K = 3.6$ (Fig.~\ref{fg:kr2}) including 
some relevant nonlinear resonances.  

\subsubsection{Pure resonance-assisted tunneling}

\begin{figure}[ht]
  \begin{center}
    \includegraphics*[width=\textwidth]{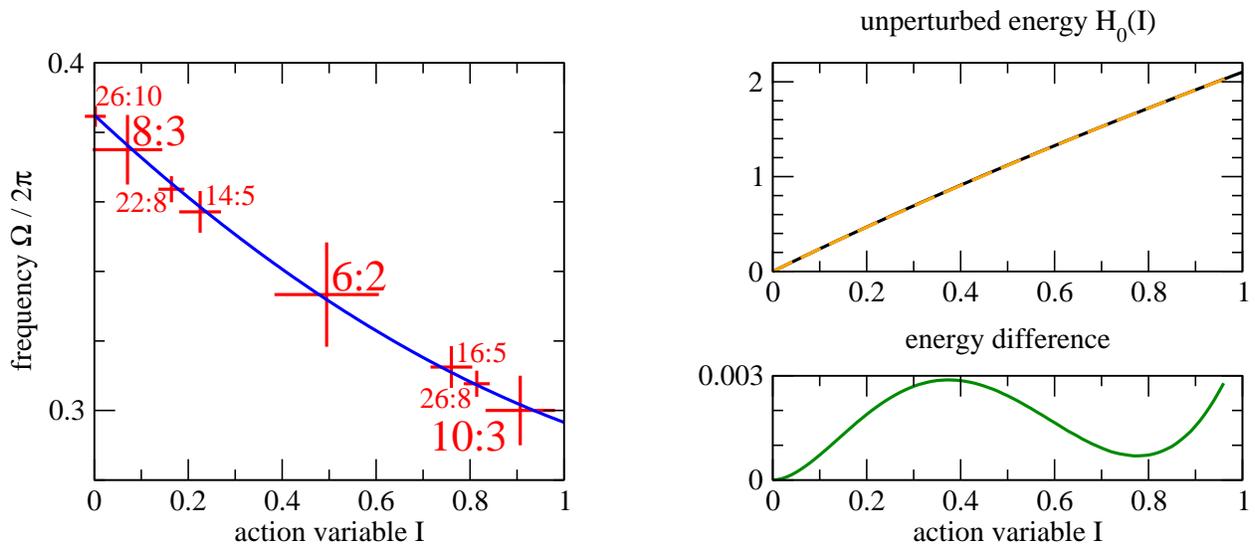}
  \end{center}
  \caption{
    Unperturbed energies and oscillation frequencies within the regular 
    island of the kicked rotor at $K=3.5$. 
    The left panel shows the action dependence of the oscillation frequencies
    as computed from a quadratic fit to individual nonlinear resonances 
    (crosses), which results in the expression
    $\Omega(I) = 2.41886 - 0.790561 I + 0.235191  I^2$ (solid line).
    The upper right panel compares the unperturbed energies resulting
    from the integration of this quadratic expression (solid line)
    with the unperturbed energies used in Ref.~\cite{LoeO10PRL} that were 
    obtained by analyzing a dense set of quasi-periodic trajectories 
    within the regular island (dashed line) \cite{thanks_steffen}.
    This latter approach yields $\Omega(I) = 2.41740 - 0.952917 I 
    + 1.00151 I^2 - 1.00153 I^3 + 0.368829 I^4$, which essentially 
    constitutes the definition of the fictitious integrable system used in 
    Ref.~\cite{LoeO10PRL}.
    Although the difference between these two approaches is rather small
    as shown in the lower right panel, it plays
    a significant role for the tunneling rates in the deep 
    semiclassical limit (see Fig.~\ref{fg:comp}). \label{fg:energies}
  }
\end{figure}

\begin{figure}[!ht]
  \begin{center}
    \includegraphics*[width=\textwidth]{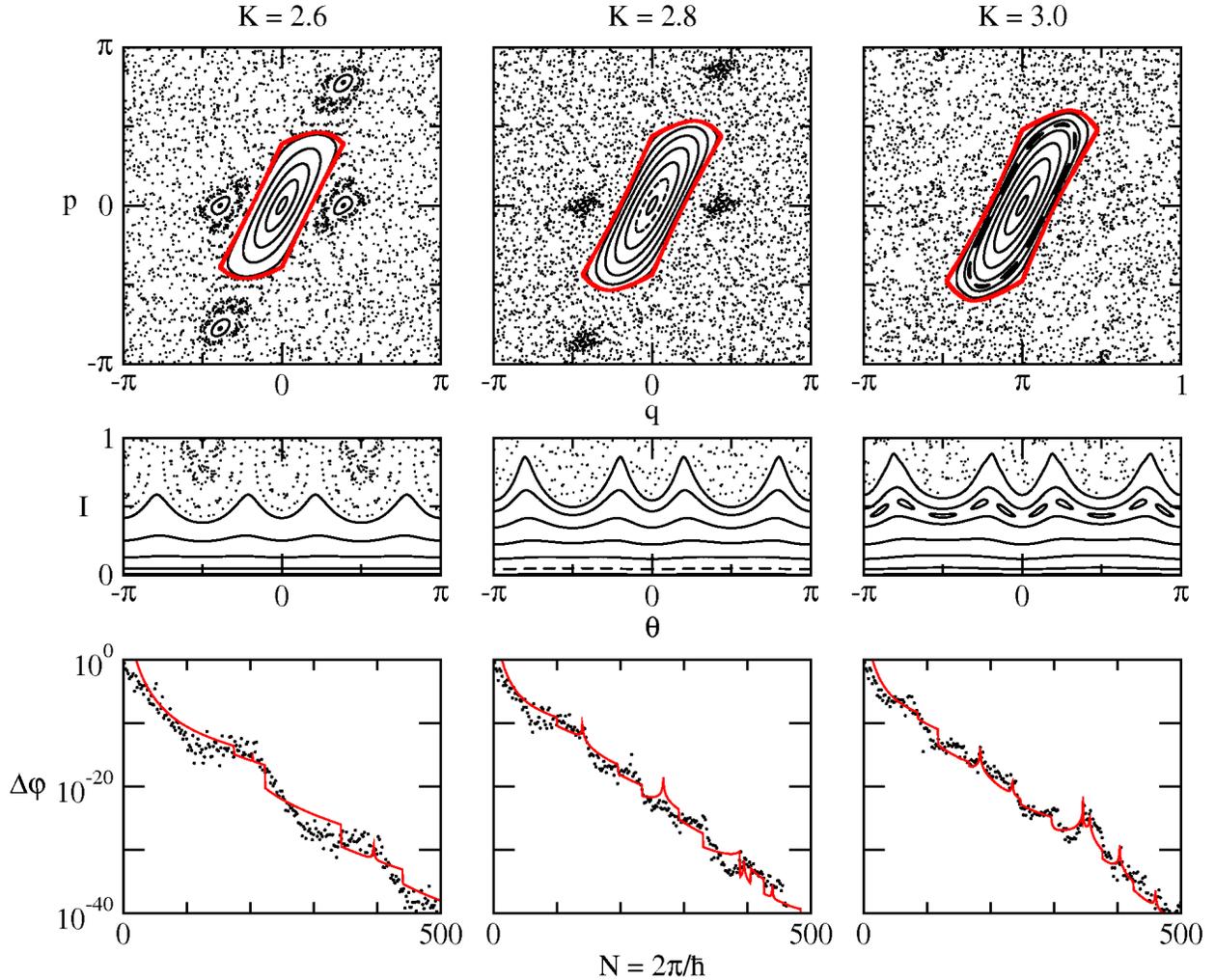}
  \end{center}
  \caption{
    Quantum and semiclassical splittings in the kicked rotor model
    for $K = 2.6$ (left column) $K=2.8$ (central column), 
    and $K=3$ (right column).
    The upper and middle panels show the classical phase space in the original
    phase space variables $p$ and $q$, with the thick curve marking the 
    effective quantum boundary of the island, and in approximate 
    action-angle variables $I$ and $\theta$.
    The lower panels display the quantum and semiclassical eigenphase 
    splittings (dots and solid lines, respectively) of the ground state 
    in the central regular island.
    For the semiclassical splittings, we used the 14:4 and 18:5 resonances
    for $K=2.6$, the 10:3 and 14:4 resonances for $K=2.8$,
    and the 10:3, 14:4, 16:5, and 22:7 resonances for $K=3$.
    As pointed out in the text, the splittings were computed with 
    a generalization of the multi-resonance expression 
    (\ref{eq:Deltaphimultires}) to mixed systems,
    using (I) the corrected action dependence
    (\ref{eq:matrelem2}) of the matrix elements and (II) unperturbed energies
    that were determined from a global parabolic fit of $\Omega(I)$,
    and (III) computing the coupling to the chaotic domain via the outermost
    nonlinear resonance, as explained at the end of section \ref{sec:tcat}.
    The discontinuous steps in the semiclassical splittings
    are actually induced by the discontinuity of the integer part in the 
    expression (\ref{eq:kc}) for the maximal number $k_c$ of couplings 
    within the island.
    \label{fg:kr1}
  }
\end{figure}

\begin{figure}[!ht]
  \begin{center}
    \includegraphics*[width=\textwidth]{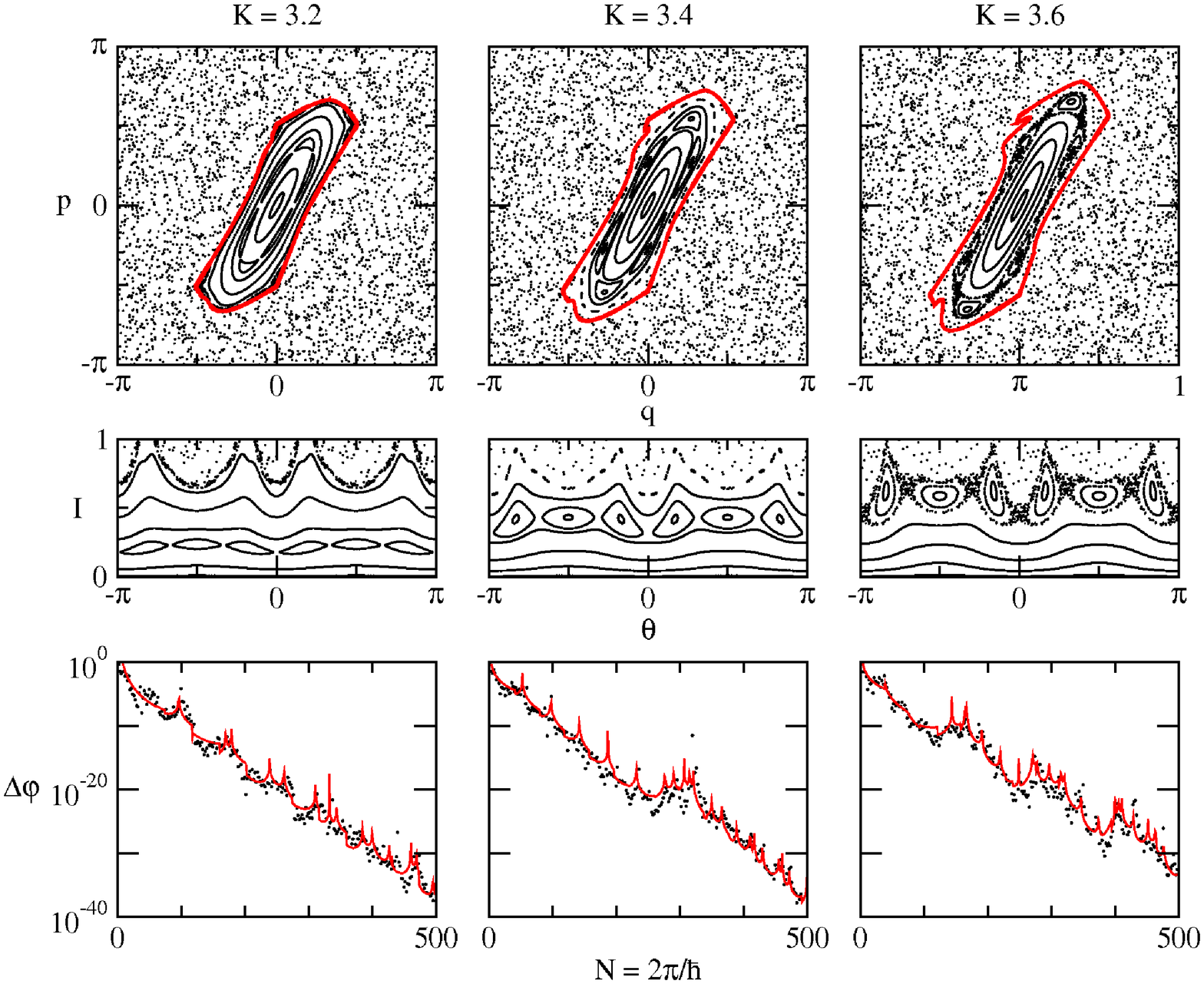}
  \end{center}
  \caption{
    Same as Fig.~\ref{fg:kr1} for $K = 3.2$, $K = 3.4$, and $K = 3.6$.
    For the semiclassical splittings, we used the 6:2 and 10:3 resonances
    for $K = 3.2$, the 6:2, 10:3, and 14:5 resonances for $K = 3.4$, and 
    the 6:2 and 8:3 resonances for $K = 3.6$.
    \label{fg:kr2}
  }
\end{figure}

We stress that the semiclassical calculations shown in
Figs.~\ref{fg:kr1} and \ref{fg:kr2} involve a few differences as
compared to some of our previous publications
\cite{EltSch05PRL,SchEltUll06,MouEltSch06PRE,WimO06PRL}. 
To start with, (I) the action dependence of the coupling coefficients 
associated with the resonances has been included 
[see Eq.~(\ref{eq:matrelem2})].  
Furthermore, the unperturbed energy differences $E_n - E_{n+kr}$ of the
quasi-modes are not computed via the quadratic pendulum approximation 
(\ref{pend}).
Instead, as illustrated on Fig.~\ref{fg:energies}, (II) a global parabolic
fit to the action dependence of the frequency $\Omega \equiv \Omega(I)$ 
was applied on the basis of the classically computed values
of the resonant actions $I_{r:s}$ and their frequencies
$\Omega(I_{r:s}) = (s/r) \, \omega$, for a sequence of
resonances with not too large $r$ and $s$.  Theses two modifications
significantly improve the reproduction of individual peak structures
in the tunneling rates.

With these improvements (I) and (II), we generally find that the quantum
splittings are quite well reproduced by our simple semiclassical
theory based on nonlinear resonances.  In particular, the location and
height of prominent plateau structures and peaks in the tunneling
rates can, in almost all cases, be quantitatively reproduced through
resonance-assisted tunneling.  The additional fluctuations of the
splittings on a small scale of $N$, however, cannot be accounted for
by our approach as they arise from the details of the eigenspectrum in
the chaotic block of the Hamiltonian. Their average size, however,
seems in good agreement with the universal prediction (\ref{eq:var})
for the variance of eigenphase splittings in chaos-assisted tunneling.

Note that there is a general tendency of the semiclassical theory 
to overestimate the exact splittings wherever the latter encounter local 
minima.
We attribute those minima to the occurrence of destructive interferences 
between different pathways that connect the ground state to a given excited 
state $|n\rangle$.
As pointed out in the discussion of Eq.~(\ref{eq:Deltaphimultires}),
such destructive interferences are not yet accounted for in our present 
implementation of resonance-assisted tunneling.
Their inclusion would require to take into account the phases $\phi_k$ 
associated with individual resonances [see Eq.~(\ref{eq:Vres})], 
the discussion of which is beyond the scope of this article.

\subsubsection{Resonance-assisted tunneling at a bifurcation}

As a last comment, we note that the influence of a nonlinear resonance
on tunneling processes in mixed systems may persist even if that
resonance is not at all manifested in the classical phase space.  This
is precisely the case at the value of the perturbation parameter at
which this resonance is bifurcating from the center of the island,
i.e., at which the central fixpoint of the island exhibits a rational
winding number corresponding to the resonance under consideration.  A
prominent example in the kicked rotor model is found at $K=2$ where
the central fix point has the winding number $0.25$. Indeed, $K=2$ is
exactly the critical value where two periodic orbits of period~4
coalesce in the center, both separately coming from the complex
phase space (their action being strictly negative for $K<2$) and
then becoming real and distinct (one stable and one unstable) for
$K>2$.

Following the normal-form arguments in Section~\ref{sec:ac}, the scaling
of the ``classical size'' of the resonance with the perturbation parameter
$K$ is, in lowest order, provided by the effective pendulum matrix element
$V_{r:s}(K) = \tilde{v}_1 [I_{r:s}(K)]^{r k / 2}$ where $I_{r:s}(K)$ 
represents the dependence of the resonant action on $K$.
However,  as can be seen in Eq.~(\ref{eq:matrelem2}), the 
associated coupling matrix elements that affect tunneling only 
depend on the prefactor $\tilde{v}_1$, which ought to be a well-behaved 
function of $K$ showing no singular behaviour at the bifurcation point.
Therefore, we have to conclude that these matrix elements remain finite
even for~$K\lesssim2$.

\begin{figure}[!ht]
  \begin{center}
    \includegraphics*[width=\textwidth]{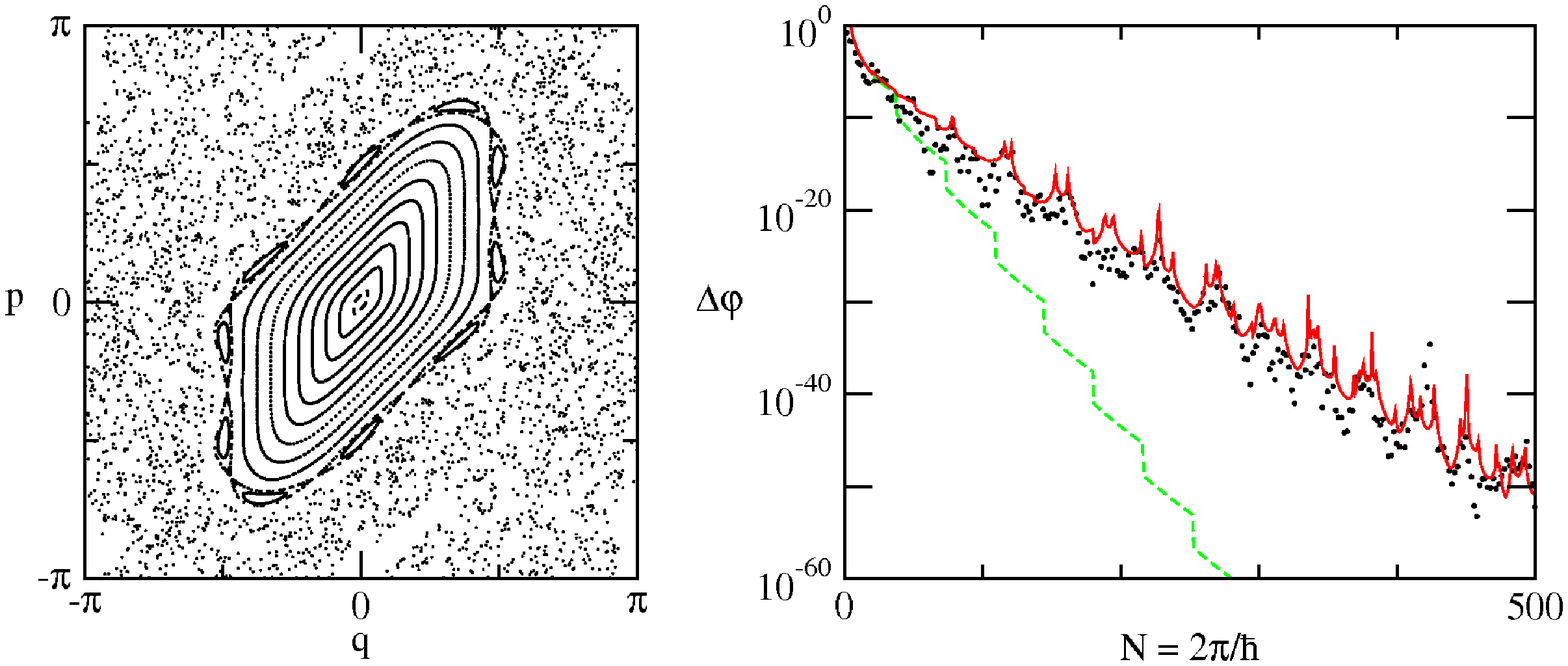}
  \end{center}
  \caption{
    \label{fg:bif}
    Resonance-assisted tunneling in the kicked rotor at a bifurcation.
    The left panel shows the classical phase space for $K = 2$,
    which contains a prominent 10:2 resonance close to the border of 
    the island, and the right panel displays the corresponding quantum 
    eigenphase splittings (dots).
    Semiclassical calculations of the splittings (solid and 
    dashed lines, using the same levels of approximation as 
    in Figs.~\ref{fg:kr1} and \ref{fg:kr2}) 
    were carried out using the 10:2 resonance only (dashed line) as well
    as a combination of the 4:1 resonance and the 10:2 resonance, the 
    former emerging at the island's center right at $K = 2$.
    The parameter $\tilde{v}_1$ associated with that 4:1 resonance
    [see Eq.~(\ref{eq:matrelem2})] was determined from the classical
    phase space at $K = 2.001$.
  }
\end{figure}

This expectation is confirmed in Fig.~\ref{fg:bif}, which shows the 
tunneling rates in the kicked rotor for $K=2$.
For this value, the only major resonance that is manifested in the
classical phase space is the 10:2 resonance whose island chain is 
located near the boundary of the main island.
Taking into account this resonance alone (as well as combining it with
other resonances of higher order, the result of which is not shown in 
Fig.~\ref{fg:bif}) apparently leads to a very strong underestimation 
of the quantum eigenphase splittings (in contrast to 
Ref.~\cite{EltSch05PRL} where the action dependence of the coupling
matrix elements was not properly incorporated).
But once we take into account the 4:1 resonance 
and compute its corresponding classical parameters
from the classical phase space at $K = 2.001$, 
we obtain a good reproduction of the quantum splittings.

\subsubsection{Combination with direct regular-to-chaotic couplings}

In the semiclassical predictions shown in Figs.~\ref{fg:kr1},
\ref{fg:kr2}, and \ref{fg:bif}, we have, following the discussion in
section~\ref{sec:RCviaNLR},  assumed that the resonance-assisted 
mechanism was providing the relevant couplings not only within the
regular island, but also (III) from the edge of the regular island to 
the chaotic sea. As pointed out in section~\ref{sec:CwCS}, the
regular-chaos boundary is, however, the place where 
approximation schemes are not controlled any longer.
It is therefore useful to compare the results
obtained in this way with those derived from the more precise
evaluation of the direct regular-chaotic couplings
(IIIb) computed with Eq.~(\ref{eq:SemiNum})
through a numerical application of the quantum evolution operator $\hat U$.

\begin{figure}[!ht]
  \begin{center}
    \includegraphics*[width=\textwidth]{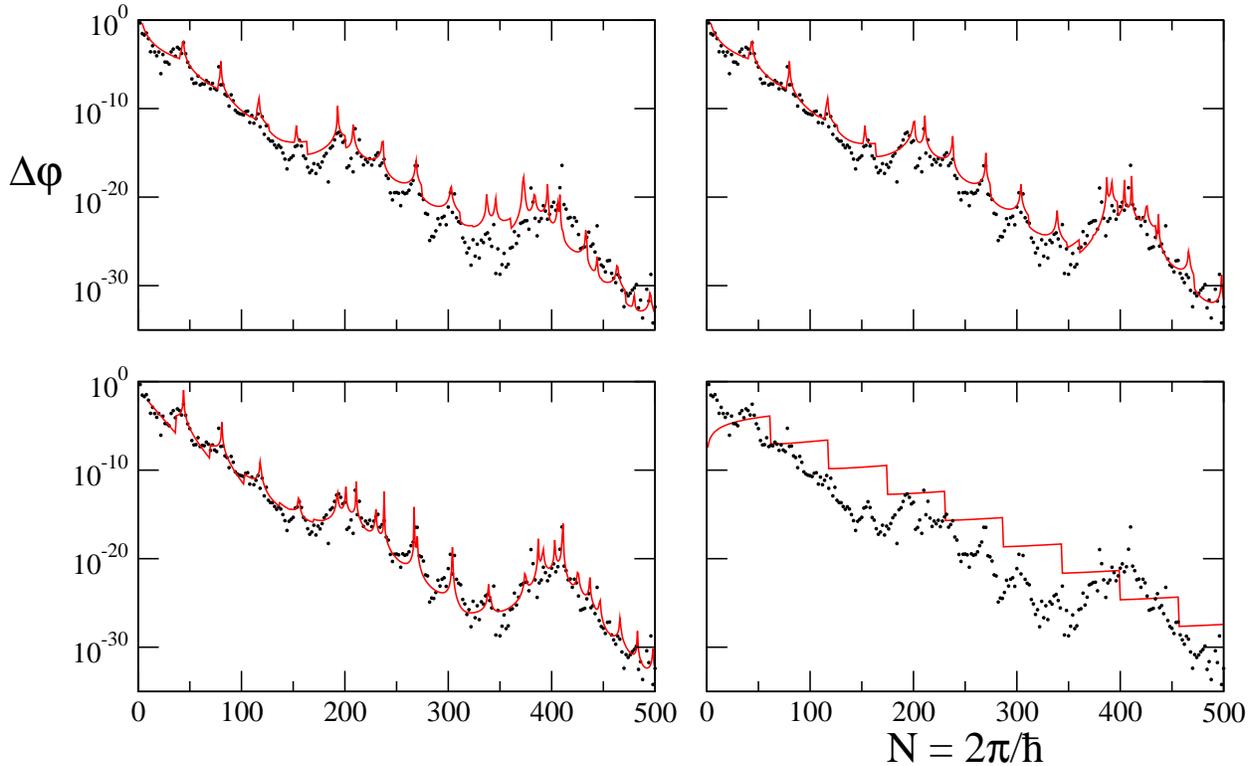}
  \end{center}
  \caption{Quantum and semiclassical splittings in the kicked rotor model
    for $K = 3.5$ (see Fig.~\ref{fg:sep} for the classical phase space).
    As in Figs.~\ref{fg:kr1} and \ref{fg:kr2}, the dots and 
    solid lines represent, respectively, the quantum splittings 
    and the semiclassical prediction. In the upper left panel, this
    prediction is based on our approach, where we take into account the
    6:2, 8:3, 10:3, and 14:5 resonances.
    The upper right panel shows the same calculation, except that a
    more refined evaluation of the unperturbed energies (and their
    action dependence) is used (IIb) (see text and Fig.~\ref{fg:energies}).
    The lower left panel shows the prediction that is obtained 
    with these improved energies and with an evaluation of the 
    direct regular-chaotic coupling (IIIb) using the fictitious 
    integrable system approach of B\"acker et al.\ \cite{BaeO08PRL}
    [which amounts to applying the quantum kicked rotor map onto the 
    corresponding eigenstate of the integrable system, see 
    Eq.~(\ref{eq:SemiNum})].
    Finally, the lower right panel shows the prediction that would be 
    obtained with the 6:2 resonance according to the approach outlined in
    Ref.~\cite{EltSch05PRL}, i.e.\ neglecting the action dependence of 
    the coupling matrix elements, neglecting the occurence of partial 
    barriers in the chaotic domain, and making a simple quadratic 
    expansion of the unperturbed energies in the vicinity of the resonance.
    \label{fg:comp}
  }
\end{figure}

Figure \ref{fg:comp} shows the resulting comparison for the eigenphase
splittings of the quantum kicked rotor at $K = 3.5$ \cite{LoeO10PRL}
(see Fig.~\ref{fg:sep} for the corresponding classical phase space).
In addition to the quantum and semiclassical splittings obtained in the
same way as in Figs.~\ref{fg:kr1} and \ref{fg:kr2}
(black and red curves, respectively), two additional
curves are shown.  The green one corresponds to a fully 
semiclassical calculation for which the regular-to-chaotic couplings
are evaluated by the resonance-assisted mechanism.
In contrast to the red curves and to the calculations in 
Figs.~\ref{fg:kr1} and \ref{fg:kr2}, however, the action dependence 
$H_0(I)$ of the unperturbed energies within the island was not obtained 
by a fit to several relevant resonances as described above, but rather
(IIb) by computing the action and the rotation number for a dense
set of trajectories within the island, from which the energies are 
deduced via the relation $\Omega(I) = dH_0(I)/dI$ \cite{thanks_steffen} 
(see, e.g., Fig.~11 of Ref.~\cite{BohTomUll93PR} and the 
associated text for a detailed discussion).  
The blue curve implements, in addition to these improved
energies, an evaluation of the direct regular-chaotic coupling  
(IIIb) using the approach of B\"acker et al.\ \cite{BaeO08PRL} 
[see Eq.~(\ref{eq:SemiNum})]; it actually corresponds to the curve 
published in Fig.~3 of Ref.~\cite{LoeO10PRL}.

What is observed in Fig.~\ref{fg:comp} is that, although there is
good agreement between the three theoretical curves and the numerical
one for a large range of $N = 2\pi/\hbar$, some significant deviations 
arise in the range $300 < N < 450$.  Remarkably, however, the most
striking discrepancies that are encountered for the red curve 
(i.e., for the full semiclassical calculation \emph{without} 
the improved energies) are essentially cured once the improved
energies are implemented (green curve).  Adding the semi-numerical
evaluation of the direct regular-to-chaotic coupling further improves 
the prediction, but to a significantly lesser degree.

Figure~\ref{fg:energies} compares $H_0(I)$ computed for the kicked rotor 
at $K=3.5$ by the two methods under consideration, namely the fitting 
approach using the most relevant resonances (which is illustrated in the 
left panel of Fig.~\ref{fg:energies}) and the more refined approach based 
on a dense set of trajectories \cite{thanks_steffen}.
Apparently, both approaches yield nearly identical energies,
with relative differences being well below one percent 
(see the lower right panel of Fig.~\ref{fg:energies}).
This underlines that small imprecisions in the prediction of the
unperturbed eigenenergies of regular quasi-modes may, under special
conditions, lead to dramatic over- or underestimations of the tunneling
rates in mixed systems.
In the particular case considered here, it seems to be the transition 
across the 8:3 resonance, located rather close to the center of the island,
which is not properly described using the more approximate energies.
This gives rise to a horizontal shift of the predicted splittings
towards smaller $N$, as is clearly seen in Fig.~\ref{fg:comp}.
This issue obviously requires further investigations.
It does, however, not seem to put into question the principal conclusion 
that the resonance-assisted coupling mechanism provides an accurate approach
to evaluate regular-to-chaotic tunneling rates in a purely semiclassical 
manner.

\section{Conclusion}

\label{sec:cc}
In summary, we have provided a comprehensive description of the 
theory of resonance-assisted tunneling in mixed regular-chaotic systems.
This description is partly based on previous publications of ours 
\cite{BroSchUll01PRL,BroSchUll02AP,EltSch05PRL,SchEltUll06,MouEltSch06PRE},
but contains also some new aspects and significant improvements especially 
concerning the determination of the matrix elements associated with
resonance-induced couplings.
Moreover, partial barriers in the chaotic domain are now incorporated into
the scheme of resonance-assisted tunneling, which generally gives rises 
to an effective enhancement of the size of the regular islands under 
consideration.
In practice, we find that the most relevant partial barriers for tunneling
are constituted by combinations of stable and unstable manifolds that are
associated with hyperbolic periodic points of low-order nonlinear resonances 
within the chaotic domain.
The application of this approach to the kicked rotor model yields rather 
good agreement between the ``exact'' eigenphase splittings of states that 
are localized in the center of the main regular island, and their 
semiclassical predictions based on nonlinear resonances.

The main message that we intend to communicate here is that
resonance-assisted tunneling not only allows one to understand the
origin of plateaus and peak structures, in the tunneling rates.  It
also provides a simple, readily implementable scheme to quantitatively
predict the appearance of such structures on the basis of purely
classical information.  In practice, one needs for this purpose to 
identify the relevant resonances in the regular phase space region 
under consideration, to find their stable and unstable fixpoints 
in the Poincar\'e surface of section, to compute their stability 
indices and the areas enclosed by their separatrix structures, 
respectively by their stable and unstable manifolds, and finally to 
compute the flux enclosed by the turnstiles in order to determine 
effective ``quantum chaos border'' of the island.  
Even for a simple model like the kicked rotor, this programme
requires much less numerical effort than a quantum calculation of the
tunneling rates.  As we demonstrated for the kicked rotor, it
provides, on the other hand, a reproduction of the quantum tunneling
rates which is extremely satisfactory from a quantitative point
of view. This numerical accuracy requires, however, a careful evaluation 
of the various classical parameters entering into the semiclassical 
calculation of the splittings. This includes the action dependence 
of the resonance-Hamiltonian couplings, the effective size of the 
regular island, and the evaluation of the unperturbed energies 
within the island.
It turns out in particular that in some circumstances, e.g. at 
$K = 3.5$ in the range $300 < N <450$, very small imprecision in the 
determination of the unperturbed eigenenergies may significantly 
affect the accuracy of the semiclassical predictions.

We expect that the framework of resonance-assisted tunneling can be 
generalized to systems with more than two effective degrees of freedom,
although the identification of important resonances might become more
involved in such systems and resonance-assisted (quantum) tunneling might
compete there with (classical) Arnol'd diffusion in the deep semiclassical 
regime \cite{Kes05PRE}. 
Another open problem which needs to be addressed in more detail concerns 
the role of nonlinear resonances in trajectory-based semiclassical
approaches to tunneling, put forward by Shudo, Ikeda, and coworkers
\cite{ShuIke95PRL,ShuIke96PRL,ShuIshIke02JPA}
(see also the corresponding chapters in this book), which involve the 
complexified classical phase space and are more rigorous from a formal
semiclassical point of view than our approach.
We strongly believe that nonlinear resonances should leave their 
characteristic traces in the self-similar complex phase space structures
that govern tunneling in this framework \cite{ShuIke95PRL,ShuIke96PRL}.
Such an insight should significantly contribute to rendering those approaches 
practicable for more complicated systems as well --- and underline the
semiclassical nature of resonance-assisted tunneling.

\section*{Acknowledgements}

We are grateful to Olivier Brodier, Dominique Delande, 
Christopher Eltschka, Kensuke Ikeda, J\'er\'emy Le Deunff, Steffen L\"ock,
Srihari Keshavamurthy, Roland Ketzmerick, Mirjam Schmid, Akira Shudo 
and Steve Tomsovic for helpful assistance and stimulating discussions.
We furthermore acknowledge financial support by the 
Deutsche Forschungsgemeinschaft (DFG) through the Forschergruppe
FOR760 ``Scattering systems with complex dynamics''.




\begin{thebibliography}{10}
\bibitem{thanks_steffen}
We are indepted to Steffen L\"ock who carried out this calculation and sent
  us the resulting data.
\end{thebibliography}
\end{document}